\documentstyle[subeqnarray,sprocl]{article}

\newcommand{\beq}{ \begin{equation} }
\newcommand{\eeq}{ \end{equation} }
\newcommand{\beqn}{ \begin{subeqnarray} }
\newcommand{\eeqn}{ \end{subeqnarray} }
\newcommand{\beqnw}{ \begin{eqnarray} }
\newcommand{\eeqnw}{ \end{eqnarray} }
\newcommand{\calle}[1]{ (\ref{#1}) }
\newcounter{exer}[section]

\newif\iffn\fnfalse

\def\centeronto#1#2{{\setbox0=\hbox{#1}\setbox1=\hbox{#2}\ifdim
\wd1>\wd0\kern.5\wd1\kern-.5\wd0\fi
\copy0\kern-.5\wd0\kern-.5\wd1\copy1\ifdim\wd0>\wd1
\kern.5\wd0\kern-.5\wd1\fi}}

\def\lspace{\ifx\answ\bigans{}\else\qquad\fi}
\def\lbspace{\ifx\answ\bigans{}\else\hskip-.2in\fi} 
\def\CA{{\cal A}}   
  \def\CG{{\cal G}} 
 \def\CJ{{\cal J}}  
   
  \def\CS{{\cal S}}

\def\bar#1{\overline{#1}}

\def\frac#1#2{{\textstyle{#1\over #2}}} 
\def\Tr{\mathop{\rm Tr}}

\def\ltap{\ \raise.3ex\hbox{$<$\kern-.75em\lower1ex\hbox{$\sim$}}\ }
\def\gtap{\ \raise.3ex\hbox{$>$\kern-.75em\lower1ex\hbox{$\sim$}}\ }
\def\gl{\ \raise.5ex\hbox{$>$}\kern-.8em\lower.5ex\hbox{$<$}\ }
\def\roughly#1{\raise.3ex\hbox{$#1$\kern-.75em\lower1ex\hbox{$\sim$}}}
\def\frac#1#2{{\textstyle{#1 \over #2}}}

\def\[{\left[}
\def\]{\right]}
\def\({\left(}
\def\){\right)}

\def\ltap{\ \raise.3ex\hbox{$<$\kern-.75em\lower1ex\hbox{$\sim$}}\ }
\def\gtap{\ \raise.3ex\hbox{$>$\kern-.75em\lower1ex\hbox{$\sim$}}\ }
\def\gl{\ \raise.5ex\hbox{$>$}\kern-.8em\lower.5ex\hbox{$<$}\ }
\def\roughly#1{\raise.3ex\hbox{$#1$\kern-.75em\lower1ex\hbox{$\sim$}}}
\def\frac#1#2{{\textstyle{#1 \over #2}}}

\def\[{\left[}
\def\]{\right]}
\def\({\left(}
\def\){\right)}

\def\Dslash{{\gamma^\mu D_\mu}}
\def\pslash{{\partial}}
\def\Cslash{{C}}
\def\Gslash{{G}}
\def\gbar{{\bar \gamma}}
\def\br{{\bf r}}

\def\IZ{\relax\ifmmode\mathchoice
{\hbox{\cmss Z\kern-.4em Z}}{\hbox{\cmss Z\kern-.4em Z}}
{\lower.9pt\hbox{\cmsss Z\kern-.4em Z}}
{\lower1.2pt\hbox{\cmsss Z\kern-.4em Z}}\else{\cmss Z\kern-.4em
Z}\fi}
\def\IA{\relax{\rm I\kern-.18em A}}
\def\IB{\relax{\rm I\kern-.18em B}}
\def\IC{{\relax\hbox{$\inbar\kern-.3em{\rm C}$}}}
\def\ID{\relax{\rm I\kern-.18em D}}
\def\IE{\relax{\rm I\kern-.18em E}}
\def\IF{\relax{\rm I\kern-.18em F}}
\def\IG{\relax\hbox{$\inbar\kern-.3em{\rm G}$}}
\def\IGa{\relax\hbox{${\rm I}\kern-.18em\Gamma$}}
\def\IH{\relax{\rm I\kern-.18em H}}
\def\II{\relax{\rm I\kern-.18em I}}
\def\IJ{\relax{\rm I\kern-.18em J}}
\def\IK{\relax{\rm I\kern-.18em K}}
\def\IL{\relax{\rm I\kern-.18em L}}
\def\IM{\relax{\rm I\kern-.18em M}}
\def\IN{\relax{\rm I\kern-.18em N}}
\def\IO{\relax{\rm I\kern-.18em O}}
\def\IP{\relax{\rm I\kern-.18em P}}
\def\IQ{\relax\hbox{$\inbar\kern-.3em{\rm Q}$}}
\def\IR{\relax{\rm I\kern-.18em R}}
\def\IW{\relax\hbox{$\inbar\kern-.3em{\rm W}$}}

\font\cmss=cmss10 \font\cmsss=cmss10 at 7pt
\def\IR{\relax{\rm I\kern-.18em R}}

\def\boxit#1{\vbox{\hrule\hbox{\vrule\kern8pt
\vbox{\hbox{\kern8pt}\hbox{\vbox{#1}}\hbox{\kern8pt}}
\kern8pt\vrule}\hrule}}
\def\mathboxit#1{\vbox{\hrule\hbox{\vrule\kern8pt\vbox{\kern8pt
\hbox{$\displaystyle #1$}\kern8pt}\kern8pt\vrule}\hrule}}

\def\unlockat{\catcode`\@=11}
\def\lockat{\catcode`\@=12}

\unlockat

\def\newsec#1{\global\advance\secno by1\message{(\the\secno. #1)}
\global\subsecno=0\global\subsubsecno=0\eqnres@t\noindent
{\bf\the\secno. #1}
\writetoca{{\secsym} {#1}}\par\nobreak\medskip\nobreak}
\global\newcount\subsecno \global\subsecno=0
\def\subsec#1{\global\advance\subsecno
by1\message{(\secsym\the\subsecno. #1)}
\ifnum\lastpenalty>9000\else\bigbreak\fi\global\subsubsecno=0
\noindent{\it\secsym\the\subsecno. #1}
\writetoca{\string\quad {\secsym\the\subsecno.} {#1}}
\par\nobreak\medskip\nobreak}
\global\newcount\subsubsecno \global\subsubsecno=0
\def\subsubsec#1{\global\advance\subsubsecno by1
\message{(\secsym\the\subsecno.\the\subsubsecno. #1)}
\ifnum\lastpenalty>9000\else\bigbreak\fi
\noindent\quad{\secsym\the\subsecno.\the\subsubsecno.}{#1}
\writetoca{\string\qquad{\secsym\the\subsecno.\the\subsubsecno.}{#1}}
\par\nobreak\medskip\nobreak}

\def\subsubseclab#1{\DefWarn#1\xdef
#1{\noexpand\hyperref{}{subsubsection}%
{\secsym\the\subsecno.\the\subsubsecno}%
{\secsym\the\subsecno.\the\subsubsecno}}%
\writedef{#1\leftbracket#1}\wrlabeL{#1=#1}}
\lockat

\def\inbar{\,\vrule height1.5ex width.4pt depth0pt}

\font\cmss=cmss10 \font\cmsss=cmss10 at 7pt
\def\IR{\relax{\rm I\kern-.18em R}}

\def\Tr{\rm Tr}
\def\Det{\rm Det}

\begin{document}

\title{TASI 2003 LECTURES ON ANOMALIES}

\author{JEFFREY A. HARVEY}

\address{Enrico Fermi Institute and Department of Physics\\
                 University of Chicago \\
                 5640 S. Ellis Avenue \\
                 Chicago, IL 60637, USA}
                 
\maketitle\abstracts{
These lecture notes review the structure of anomalies and present some of their applications in field theory, string theory
and M theory. They expand on material presented at the TASI 2003 summer school and the 2005 International Spring School on String Theory in
Hangzhou, China.
}

\vfil \eject

\tableofcontents

\vfil \eject

\setcounter{section}{-1}

\section{Introduction}

The study of anomalies has many applications in field theory and string theory. These
range from phenomenological applications such as the calculation of the decay rate for neutral pions
into two photons, the computation of quantum numbers in the Skyrme model of hadrons, and
mechanisms for baryogenesis in the Standard Model;  to more abstract applications such as the
study of dualities in gauge theory, the computation of anomalous couplings in the effective theory
of D-branes, and the analysis of Black Hole entropy. 

Anomalies are often a useful first line of attack in trying to understand new systems. This is because
the presence of anomalies, or the way they are canceled, can often be studied without knowing
the detailed dynamics of the theory. They are in a way topological properties of the
theory and thus can be studied by approximate methods.

The basic definition of an anomaly is the following.  Consider a quantum theory which has a symmetry group $G$ which leaves the classical action
invariant  ($\delta S_{cl}=0$).  We say that $G$ is anomalous if $G$ is violated
in the full quantum theory. Thus anomalous symmetries are symmetries of classical theories
which do not survive the transition to quantum mechanics. 
The method of computation and physical 
implications of anomalies depend on the structure of $G$. In particular,
$G$ can be either discrete or continuous, and it can be either a global
symmetry or a gauge (local) symmetry. 

If the symmetry $G$ is a global symmetry then anomalies in $G$ do not
indicate any inconsistency of the theory,  but they  often have interesting
physical consequences. Historically the most important example of this
type is the anomaly in the axial current which is important in understanding
the decay rate for $\pi^0 \rightarrow \gamma \gamma$. This is the subject of lecture 1.

Classically, invariance under a continuous global symmetry group $G$ implies the existence
of conserved currents $j_\mu^a$ with $a$ labelling the generators of $G$: 
$\partial^\mu j_\mu^a = 0$. If the symmetry is anomalous then
there are quantum corrections which make the divergence of $j_\mu^a$ non-zero,
$\partial^\mu j_\mu^a =  {\CA}^a(\phi,\partial \phi)$. Thus the variation of the quantum effective action
under a symmetry transformation labelled by $v^a$ is
\beq
\label{onea}
\delta_{v} S_{\rm eff} = \int v^a \partial^\mu j_\mu^a = \int v^a  {\CA}^a
\eeq
and the quantity ${\CA}^a$ is the anomaly. 
Anomalies can be more precisely phrased as violations of the Ward identities following from
$G$ invariance, a point of view which will be developed further in sec 3.1.

On the other hand, if $G$ is a gauge symmetry then anomalies indicate
a fundamental inconsistency of the theory and must vanish. Recall
that gauge symmetries are not symmetries in the conventional sense of
symmetries that act on the configuration space and lead to identical
physics. Rather, they are redundancies in our description of the physics
when we work in the space of gauge fields rather than its quotient by gauge
transformations. Anomalies in a redundancy would not be a
good thing. More concretely, when we work in the space of gauge field configurations, we need gauge
invariance to remove negative norm  states from the spectrum and a lack of
gauge invariance due to 
anomalies would lead to fundamental inconsistencies \cite{GrossPV}.

We can also distinguish between several possible types of gauge anomaly. Continuous gauge
transformations can either be local, here meaning that they can be continuously connected
to the identity transformation, or global, meaning that they cannot be so connected. 
It is possible for a gauge theory to have anomalies in global but not local gauge
transformations. A famous example is $SU(2)$ gauge theory in $D=4$ spacetime dimensions
with a single Weyl fermion in the two-dimensional representation of $SU(2)$. Working
in Euclidean space, the fact that $g(x) \rightarrow 1$ as $|x| \rightarrow \infty$ means that we can
identify the boundary of $R^4$ with a single point, so that spacetime can be effectively viewed
as $S^4$.  Then gauge transformations are maps from $S^4$ to $SU(2)$, and such maps are classified by
$\pi_4(SU(2))=Z_2$. Thus there are gauge transformations which are not connected to
the identity. It was shown in \cite{wittglob} that the quantum effective action is not invariant
under global $SU(2)$ transformations, that is, there is an anomaly in global $SU(2)$ gauge
transformations. This inconsistency of the theory came as a relief,
since otherwise one would have had to make sense of other rather peculiar properties of
this theory such as the odd number of fermion zero modes in an instanton background. 

One can also consider theories where gauge symmetries are discrete and ask if these
discrete transformations are symmetries of the quantum theory. See for example
\cite{kw,banksdine,PreskillBM,BanksAG,IbanezHV,cm}.

Gravity is also a gauge theory of a sort. When fermions are incorporated into gravitational
theories one requires both diffeomorphism symmetry and local Lorentz symmetry. These symmetries can also be anomalous, although they are somewhat more exotic than
gauge anomalies  in that
gravitational anomalies only appear in spacetime dimensions $D=2+4k$ with
$k$ integer. A seemingly even more
exotic possibility is that a theory could have global gravitational anomalies when the group
of diffeomorphisms has components not connected to the identity \cite{wittgg}. A simple and important
example of this occurs in string theory when one studies one-loop diagrams in string
perturbation theory. The string world-sheet is then a  two-torus, $T^2$, and
the group of global diffeomorphisms is
$SL(2,\IZ)$.  Invariance under $SL(2,\IZ)$ is also known as modular invariance and provides an important constraint on the structure of chiral string theories \cite{wittgg,ghmr}.

A final class of anomalies of central importance in particle physics are the trace, scale, or
conformal anomalies. These occur in theories which are classically scale and/or conformally
invariant, but where the invariance is broken by quantum effects. Unlike the other anomalies
discussed here, which only occur in certain specific theories, anomalies in scale invariance are
generic due to the non-trivial renormalization group flow of interacting quantum field
theories.  It is only in very special theories like $N=4$ Super Yang-Mills or special
theories with $N=2$ or $N=1$ supersymmetry in four dimensions that they are absent. 

In these lectures I will first introduce some of the basic ideas in simple systems. I will then discuss some
four-dimensional, ``real world" applications, present some of the mathematical tools needed
to do computations with anomalies in higher-dimensional theories, and then end with some
applications of anomalies to branes in string theory and M-theory. Although I will cover a number
of topics in anomalies, the central topic will be anomaly inflow and its applications. There a
number of  good reviews of anomalies which focus on other topics.  For a brief overview of anomalies see the review by Adler \cite{AdlerIH}. The review of Alvarez-Gaume
and Ginsparg \cite{agg} has a comprehensive discussion of the topological interpretation of anomalies
as well as many useful formulae.  Scrucca and Serone \cite{ScruccaJN} focus on
anomalies in theories with extra dimensions and have an extensive treatment of anomalies in
orbifold theories. The Bilal and Metzger review \cite{BilalES} discusses all aspects of anomaly
cancellation in M theory, including fivebrane anomalies and anomaly cancellation in heterotic
M theory.  The Green-Schwarz mechanism of anomaly cancellation in superstring theory
is discussed in the textbooks \cite{gsw} and \cite{Poltwo}. In preparing these lectures I found the treatment of anomalies in the textbooks by
Peskin and Schroeder \cite{ps} and by Weinberg \cite{weinberg} to be particularly useful.
This review will not discuss anomalies in discrete symmetries, whether global or
gauged.  Nor will it
discuss scale anomalies or anomalies in supersymmetric theories, except in passing. These are
all interesting topics but would take us too far afield.

\subsection{Conventions}

We will sometimes write gauge fields  in components as $A_\mu^i$ with
$\mu$ a spacetime vector index and $i$ a gauge index.  In later sections we write gauge fields as 
$1$-forms taking values in the Lie algebra
$\CG$
of a compact Lie group $G$. In the earlier ``phenomenological'' sections we
use Hermitian generators of $G$,  while in more mathematical sections we choose anti-Hermitian matrices
$(\lambda^i)^a_b$ which span the adjoint representation of
$\CG$ and write $A= A_\mu^i \lambda^i dx^\mu$.  The representation matrices
are normalized so that
$$ {\rm Tr }\lambda^i \lambda^j = {1 \over 2} \delta^{ij}. $$
The
covariant derivative is $D = d + [A, ~~]$ and the curvature
is  $F = dA + A^2$.  Since we wish to couple fermions to gravity
we will decompose the metric in terms of vielbeins, $g_{\mu \nu} =
\eta_{ab}e^a_\mu e^b_\nu$ with Greek indices used for coordinate
frame indices and Latin letters for tangent space indices. 

The gamma matrices $\gamma^a$  obey $\{ \gamma^a,\gamma^b \} = 2 \eta^{ab}$.
In $2n$ spacetime dimensions, spinors can be divided into positive and negative chirality
components according to whether they have eigenvalue $\pm 1$ with respect to the
generalization of $\gamma^5$ in four dimensions to $2n$ dimensions:
$$ \gbar = \eta \prod_{a=1}^{2n} \gamma^a $$
where $\eta$ is a phase, equal to $i^n$ in Euclidean space and $i^{n-1}$ in
Minkowski space. Our metric convention is $\eta^{ab} = \rm{diag}(1,-1, \cdots  -1).$

We will often encounter differential forms which represent characteristic classes and
a set of differential forms related to these by the ``descent procedure" which will be
discussed in these lectures. For these forms we use a notation where subscripts indicate
the degree of the form and superscripts in parentheses indicate the order of the form
in the parameter of the gauge variation. Thus $\omega_2^{(1)}$  denotes a $2$-form which is
linear in the parameter of the gauge variation.  We will also encounter formal sums of differential
forms of different degrees. If $\alpha$ is such a sum, we will indicate the $n$-form part of
$\alpha$ by $\alpha|_n$. 

Finally, while I have attempted to get factors of $2$ and $\pi$ correct, no serious attempt
has been made in these notes to check signs.  For anomaly cancellation in M-theory these
have been worked out carefully in the review \cite{BilalES}.

\subsection{Exercises}
Each lecture is followed by a few exercises which should be attempted by students
wanting to have a full grasp of the material.

\section{Lecture 1: The Chiral anomaly}

\subsection{ $\pi^0 \rightarrow \gamma \gamma$}

After these generalities let me go back to the beginning. The story of anomalies could be said to
originate in a computation of the rate for the decay 
\beq
\label{zeroa}
\pi^0 \rightarrow \gamma \gamma.
\eeq
Since the $\pi^0$ is electrically neutral, it doesn't couple directly to electromagnetism.  There can 
however be a coupling induced at the one-loop level. The lowest dimension parity invariant operator one can write down which would lead to such a decay process is
\beq
\label{zerob}
{\cal L}_{\pi \gamma \gamma}= A \pi^0 \epsilon^{\mu \nu \lambda \rho} F_{\mu \nu} F_{\lambda \rho} .
\eeq

A pre-QCD computation (in 1949!) by Steinberger \cite{SteinbergerWX} (see also
\cite{fukuda})  used the coupling of pions to the nucleon doublet $N$ of the form
\beq
\label{zeroc}
G_{\pi N} \vec \pi \cdot \bar N \gamma^5 \vec \sigma N
\eeq
to compute the one-loop diagram with virtual nucleons running in the loop and
one external pion and two external photons  and obtained a result
which is equivalent to \calle{zerob} with 
\beq
\label{zerod}
A=e^2 G_{\pi N} / 32 \pi^2 m_N
\eeq
with $m_N$
the nucleon mass. This leads to a decay rate which agrees to within factors of a few
with the experimental value $\Gamma \sim 10^{16} {\rm sec}^{-1}$.

However it was later realized by Nambu that the pion should be thought of
as a Nambu-Goldstone boson resulting from the spontaneous breaking of chiral
symmetry by the QCD vacuum, $SU(2)_L \times SU(2)_R \rightarrow SU(2)_V$.
This implies that
pions only have derivative couplings up to terms suppressed by powers of
$m_\pi^2/m_N^2$.
But in this case, the coefficient $A$ should be suppressed 
by $m_\pi^2/m_N^2$ relative to the ``naive'' value \calle{zerod}  (the Sutherland-Veltman theorem
uses PCAC to show that the matrix element for $\pi^0 \rightarrow \gamma \gamma$ vanishes in
the soft pion limit \cite{Suthvelt} ).
This however is inconsistent with experiment. 
Thus there must be something ``anomalous'' going on that invalidates this reasoning.  
The anomalous behavior was understood in 1969 due to the work of Adler \cite{Adler} and 
Bell and Jackiw \cite{Bell}.
What they found is that there is a quantum violation of part of the 
$SU(2)_L \times SU(2)_R$ symmetry in the presence of electromagnetism which is 
independent of the quark masses. We will discuss their result (in modern language)
in a following section, but first we turn to a simpler system where the anomaly  in a chiral
symmetry can be
computed in a particularly straightforward manner.

\subsection{The axial current anomaly in $1+1$ dimensions}

For our first example we consider QED in two spacetime dimensions
described by the Lagrangian
\beq
\label{onee}
{\cal L} = -{1 \over 4 e^2} F_{\mu \nu} F^{\mu \nu} +
i \bar \psi \gamma^\mu D_\mu \psi 
\eeq
where
\beq
\label{onef}
\psi = \pmatrix{\psi_+ \cr \psi_- \cr} 
\eeq
is a two-component Dirac spinor and we  choose a basis of
gamma matrices so that
\beq
\label{oneg}
\gamma^0 = \pmatrix{0 & -i \cr
                             i & 0 \cr}, \qquad
         \gamma^1 = \pmatrix{0 & i \cr
                            i & 0 \cr}, \qquad
       \gbar = \pmatrix{1 & 0 \cr
                             0 & -1 \cr}.
 \eeq
Here $\gbar = \gamma^0 \gamma^1$ is the analog of $\gamma^5$ in four dimensions,
that is $(1 \pm  \gbar )/2$ are the projection operators onto chiral representations of the
Lie algebra of the Lorentz group.

Classically this theory is invariant under the vector transformation
$\psi \rightarrow e^{i \alpha} \psi$ and the chiral
transformation $\psi \rightarrow e^{i \beta \gbar} \psi$, leading
to conservation of the vector and axial currents
\beqn
\label{oneh}
j_\mu^V & = &  \bar \psi \gamma_\mu \psi \\
                   j_\mu^A & =  & \bar \psi \gamma_\mu \gbar  \psi  
\eeqn
Note that $\gamma^\mu \gbar = - \epsilon^{\mu \nu} \gamma_\nu$ so that
the two currents are related by $j_\mu^A = - \epsilon^{\mu \nu} j_\nu^V$.

We can discover an anomaly in the conservation of $j_\mu^A$ by studying
the one-loop correction to the vector current in the presence of a background
gauge field $A_\mu$. In momentum space this is given by the one-loop
diagram with one external current insertion and one insertion of the background
gauge field:
\beq
\label{onei}
\langle j^{ V \mu}(q) \rangle_A =  \Pi^{\mu \nu} A_\nu(q)
\eeq
where $\Pi^{\mu \nu}$ is the usual one-loop vacuum polarization diagram.  Computing
this diagram so as to maintain gauge invariance (e.g. by dimensional
regularization) leads to
\beq
\label{onej}
i \Pi^{\mu \nu}(q) = i (q^2 \eta^{\mu \nu} - q^\mu q^\nu)
\Pi(q^2) 
\eeq
with 
\beq
\label{onejz}
\Pi(q^2) = {1 \over \pi q^2}.
\eeq

The pole in $\Pi$ at $q^2=0$ implies that the photon has acquired a mass
$m_{\gamma}^2 = e^2/\pi$ as can be seen by summing the geometric series for
the photon propagator.
In spite of this, the vector gauge current is still
conserved, $q_\mu \langle j^{V \mu}(q) \rangle_A = 0$, but computing
the divergence of the axial current gives
\beq
\label{onek}
q_\mu \langle j^{A \mu}(q) \rangle_A = - q_\mu
\epsilon^{\mu \nu} \langle j_\nu^V \rangle_A = {1 \over \pi}
\epsilon^{\mu \nu}q_\mu A_\nu ,
\eeq
which in position space implies
\beq
\label{onel}
\partial_\mu j^{A \mu} = {1 \over 2 \pi} \epsilon^{\mu \nu}
F_{\mu \nu} . 
\eeq

So, we have found an anomaly, but it is natural to ask whether the anomaly we have
found depends on the choice of regulator. What would have happened if we had used
Pauli-Villars, or a momentum space cutoff, or some other method to regularize the theory?
Following the discussion in \cite{ps} we can reason as follows. Whatever choice of regulator
we make, dimensional analysis and Lorentz invariance tells us that the vacuum polarization
will take the form
\beq
\label{onemm}
i \Pi^{\mu \nu}(q) = i (A(q^2)  \eta^{\mu \nu} - B(q^2) {q^\mu q^\nu \over q^2}).
\eeq
Computing the divergence of the vector and axial currents we find
\beq
\label{onenn}
q_\mu \langle j^{\mu V} (q) \rangle_A = - q^\mu A_\mu(q) (A(q^2) - B(q^2))
\eeq
and
\beq
\label{oneoo}
q_\mu \langle j^{\mu A}(q) \rangle_A = A(q^2)  \epsilon^{\mu \nu} q_\mu A_\nu(q).
\eeq

Now, a little thought or calculation shows that $A(q^2)$ is logarithmically divergent, so
its value certainly depends on the choice of regulator.  $B(q^2)$ on the other hand, is
finite, independent of the choice of regulator, and in fact determined entirely by the
infrared behavior of the theory since it is the residue of the pole at  $q^2=0$. Now we could
certainly regularize the theory so that $A(q^2)=0$, but then \calle{onen} shows that
the vector current would have a non-zero divergence which would violate gauge invariance.
Gauge invariance requires that we regularize the theory so that $A(q^2)=B(q^2)$,
and since $B(q^2)$ is non-zero, we are then forced into having an anomaly in the
divergence of the axial current, no matter what choice or regulator we use. This example
shows the intricate interplay between UV divergences, IR behavior, and gauge invariance
which is characteristic of anomalies.

\subsection{Fujikawa analysis of chiral anomalies}

The
calculation in the previous section of the chiral anomaly in $1+1$ dimensions can be generalized to Dirac fermions
in $2n$ spacetime dimensions. There are many equivalent ways to compute
the anomaly. These include:
\begin{itemize}
\item  A direct calculation of  the $(n+1)$-gon diagram with one insertion of
$j^A_\mu$ and $n$ insertions of the background gauge field using a regulator which
maintains gauge invariance.
\item Point Splitting: The current involves the product of field
operators and so is potentially divergent. To regulate this split the
fermion fields apart and insert a Wilson line to maintain gauge
invariance. Thus one defines the current as
\beq
\label{onem}
j^A_\mu = \lim_{\epsilon \rightarrow 0} \bar \psi(x+\epsilon/2)
\gamma_\mu \gbar e^{-ie\int_{x-\epsilon/2}^{x+\epsilon/2} A}
\psi(x-\epsilon/2)
\eeq
and computes the divergence as $\epsilon$ is taken to zero.
\item The Fujikawa method \cite{fuji}. A careful definition of the measure $ D \bar \psi
D \psi$ in the path integral is given in terms of the spectrum of
the Dirac operator and then one finds that the measure is not
invariant under chiral transformations.
\end{itemize}

All these methods lead to the same conclusion, in $2n$ spacetime dimensions, there is a one-loop
anomaly in the divergence of the axial current given by
\beq
\label{onen}
\partial_\mu j^{A \mu} = {2 (-1)^{n+1}  \over n! (4 \pi)^n}
\epsilon^{\mu_1 \cdots \mu_{2n}}  F_{\mu_1 \mu_2} \cdots F_{\mu_{2n-1}
\mu_{2n}}
\eeq

Here I will follow the last approach pioneered by Fujikawa.  His approach is conceptually attractive and computationally powerful. To explain the method I will first consider a simple example in detail and then summarize some generalizations.
We start with a charged massless fermion $\psi$ coupled to electromagnetism in $3+1$ dimensions. The partition function is given by
\beq
\label{zeroe}
Z= \int DA_\mu D\psi D \bar \psi e^{iS[A,\psi,\bar \psi]}
\eeq
with classical action
\beq
\label{zerof}
S= \int d^4 x  \left( -{1 \over 4 e^2} F_{\mu \nu}F^{\mu \nu} + \bar \psi i \Dslash \psi  \right) .
\eeq
This theory has a chiral symmetry
\beq
\label{zerog}
\psi \rightarrow e^{i \alpha \gbar} \psi = \psi + i \alpha \gbar \psi + \cdots
\eeq
with a corresponding Noether current $j_\mu^A= \bar \psi \gamma_\mu \gbar \psi$.

The classical conservation law $\partial^\mu j_\mu^A=0$ is in the quantum theory replaced by
Ward identities.  The standard path integral derivation of Ward identities goes as follows.
We consider the change of variables in the path integral
\beqn
\label{zeroh}
\psi(x) & \rightarrow & \psi'(x)  =  \psi(x) + \epsilon(x), \\
                                     \bar \psi(x) & \rightarrow & \bar \psi'(x)  =  \bar \psi(x) + \bar \epsilon(x) . 
\eeqn
According to the standard rules of path integration, this change of variables should leave the
path integral unchanged so that
\beq
\label{zeroi}
\int D \psi D \bar \psi e^{i \int d^4x {\cal L}[\psi, \bar \psi]} =
\int D \psi' D \bar \psi' e^{i \int d^4x {\cal L}[\psi', \bar \psi']}  .
\eeq
We now apply this with $\epsilon(x) = i \alpha(x) \gbar \psi(x)$ and
$\bar \epsilon(x) = \bar \psi i \alpha(x) \gbar$. This corresponds to an infinitesimal
chiral transformation, but with a spacetime dependent parameter $\alpha(x)$. 

For $\alpha$ constant the Lagrangian  ${\cal L}$ is invariant, so the first order change in
${\cal L}$ must be proportional to $\partial_\mu \alpha$ and indeed one easily finds that
\beq
\label{zeroj}
\int d^4x \left( \bar \psi' i \Dslash \psi'  \right) = \int d^4x \left( \bar \psi i \Dslash \psi - \partial_\mu \alpha  \bar \psi \gamma^\mu \gbar \psi  \right) .
\eeq
If we then assume that the measure is invariant, $D \psi' D \bar \psi' = D \psi D \bar \psi$,
(an assumption which we will soon see is invalid) then integrating by parts and varying with respect to $\alpha$ gives the Ward identity
\beq
\label{zerok}
\partial_\mu \langle \bar \psi \gamma^\mu \gbar \psi \rangle =0.
\eeq

Fujikawa pointed out that the assumption that the measure is invariant is not necessarily
valid. This provides a very nice way of understanding why a classical symmetry might fail to be a symmetry of the quantum theory. To go from a classical theory to its quantum counterpart in the path integral formalism we need not only the classical action, but also a measure in the path integral. If the measure is not invariant then the quantum theory will not  inherit the classical symmetries of the action. We can check this idea in our specific example by giving a more precise definition of the measure
and then checking invariance under chiral transformations.

To do this we will expand $\psi $ in terms of  orthonormal eigenstates of $i \Dslash$:
\beq
\label{zerol}
i \Dslash \phi_m = \lambda_m \phi_m, \qquad \bar \phi_m i \overleftarrow{\Dslash} = \lambda_m \bar \phi_m,
\eeq
and expand
\beq
\label{zerom}
\psi(x) = \sum_m a_m \phi_m(x), \qquad \bar \psi(x) = \sum_m \bar a_m \bar \phi_m(x),
\eeq
where $a_m, \bar a_m $ are Grassmann variables multiplying the c-number eigenfunctions. We then define the measure by
\beq
\label{zeron}
D \psi D \bar \psi = \prod_m da_m d \bar a_m.
\eeq

We now make the change of variables as before and find that this induces a change in the
coefficients $a_m$:
\beq
\label{zeroo}
a'_m = \int d^4x \phi_m(x) ^\dagger \sum_n(1+i \alpha(x) \gbar )\phi_n(x) a_n
\eeq
which we will write in short-hand notation as
\beq
\label{zerop}
a'_m = \sum_n (\delta_{mn} + C_{mn}) a_n 
\eeq
with
\beq
\label{zeroq}
C_{mn}=i \int d^4x \phi_m^\dagger \alpha \gbar \phi_n.
\eeq
We can then compute the change in the measure from the Jacobian of this transformation,
and taking into account the Grassmann property of the $a_m$ find
\beq
\label{zeror}
D \psi' D \bar \psi' = ({\rm det}(1+C))^{-2} D \psi D \bar \psi .
\eeq

Thus to compute the change in the measure we need to compute ${\rm det}(1+C)$. Working
to first order in $\alpha$ and hence to first order in $C$ we have
\beq
\label{zeros}
{\rm det}(1+C)=e^{\Tr \ln(1+C)}=e^{\Tr C + \cdots} 
\eeq
so that to this order
\beq
\label{zerot}
{\rm det}(1+C)^{-2}=e^{-2 i \int d^4x \alpha(x) \sum_n \phi_n^\dagger(x) \gbar \phi_n(x)}
\eeq

Formally, the coefficient of $\alpha(x)$ in the exponent of \calle{zerot} is $\Tr \gbar$. This
trace includes a trace over the Lorentz indices of $\gbar$, which of course gives
zero, but also a trace over the infinite number of eigenstates of $i \Dslash$, which gives
infinity. In other words, \calle{zerot} is not defined without some regularization scheme. To regulate
\calle{zerot} we will define momentum integrals by continuation to Euclidean space and regularize
the sum by cutting off the sum at large eigenvalues via
\beq
\label{zerou}
\sum_n \phi_n^\dagger(x) \gbar \phi_n(x) \equiv \lim_{M \rightarrow \infty} \sum_n \phi_n^\dagger(x) \gbar \phi_n(x) e^{\lambda_n^2/M^2}.
\eeq
The sign of the last exponential in \calle{zerou} may look wrong, but we will see in a moment that
it is what we need to regulate the sum when we continue to Euclidean space.

Since the $\phi_n$ are eigenfunctions of $i \Dslash$ we can also write \calle{zerou} as
\beq
\label{zerov}
\lim_{M \rightarrow \infty} \sum_n \phi_n^\dagger \gbar e^{(i \Dslash)^2/M^2} \phi_n = \lim_{M \rightarrow \infty} \langle x| tr[\gbar e^{(i \Dslash)^2/M^2}]|x \rangle.
\eeq
Now $(i \Dslash)^2 = -D^2 +(1/2) \sigma^{\mu \nu}F_{\mu \nu}$ with 
$\sigma^{\mu \nu}= (i/2)[\gamma^\mu,\gamma^\nu]$ so we are left with the evaluation of
\beq
\label{zerow}
\lim_{M \rightarrow \infty} \langle x| tr[\gbar e^{(-D^2+(1/2)\sigma^{\mu \nu}
F_{\mu \nu})/M^2]}|x \rangle .
\eeq

We now need to figure out what terms contribute in the limit that $M$ goes to infinity. 
We can expand in powers of the background electromagnetic field, writing
$-D^2= -\partial^2 + \cdots$. Then the term with no powers of the background field
involves the integral (after continuing to Euclidean space)
\beq
\label{zerox}
\langle x|e^{-\partial^2/M^2}|x \rangle = i \int {d^4 k_E \over (2 \pi)^4} e^{-k_E^2/M^2}
= {i M^4 \over 16 \pi^2}.
\eeq
However, the trace of $\gbar$ vanishes so there is no contribution which is independent
of the background field. Bringing down one power of the background field also vanishes
since $\Tr \gbar \sigma^{\mu \nu}=0$. Terms with more than two powers of the
background field vanish in the limit $M \rightarrow \infty$. Thus we are left with a single term which is finite and non-zero in the limit $M \rightarrow \infty$ which results from expanding to second 
order in the background gauge field: 
\beq
\label{zeroy}
\lim_{M \rightarrow \infty} \Tr \left[\gbar {1 \over 2} \left({1 \over 2 M^2} \sigma^{\mu \nu}
F_{\mu \nu} \right)^2 \right ] \langle x|e^{-\partial^2/M^2} |x \rangle = - {1 \over 32 \pi^2} \epsilon^{\alpha \beta \mu \nu} F_{\alpha \beta}F_{\mu \nu}.
\eeq

We thus have
\beq
\label{zeroz}
{\rm det}(1+C)^{-2} = e^{i \int d^4x \alpha(x) \left( {1 \over 16 \pi^2}\epsilon^{\alpha \beta \mu \nu} F_{\alpha \beta}F_{\mu \nu} \right)}
\eeq
and the change of variables we used in the proof of the Ward identity thus leads to the
partition function
\beq
\label{zeroaa}
Z[A]= \int D \psi D \bar \psi e^{i \int d^4x \left( \bar \psi i \Dslash \psi + \alpha(x)
\left(\partial^\mu j_\mu^A + (1/ 16 \pi^2) \epsilon^{\alpha \beta \mu \nu} F_{\alpha \beta} F_{\mu \nu}\right) \right)}
\eeq
which after varying with respect to $\alpha$ gives the Adler-Bell-Jackiw anomaly
\beq
\label{zerobb}
\partial^\mu j_\mu^A = - {1  \over 8 \pi^2} {\tilde F}^{\mu \nu} F_{\mu \nu}
\eeq
where we have defined
\beq
\label{zerobc}
{\tilde F}^{\mu \nu} = {1 \over 2} \epsilon^{\mu \nu \alpha \beta} F_{\alpha \beta}
\eeq
and have removed a factor of $i$ in the final answer to express the answer in Minkowski space.

\subsection{Gravitational contribution to the chiral anomaly}

Fujikawa's method can also be applied to compute the anomaly in the chiral current when
fermions are coupled to gravity \cite{fuji}. The derivation proceeds along the same line as for
the gauge anomaly with some minor technical complications and leads to a gravitational contribution
to the divergence of the axial current given by
\beq
\label{fujigrav}
D^\mu j_\mu^A = - {1 \over 384 \pi^2} {1 \over 2} \epsilon^{\mu \nu \alpha \beta}
R_{\mu \nu \sigma \tau} {R_{\alpha \beta}}^{\sigma \tau} .
\eeq

\subsection{Why anomalies are one-loop exact}

We have so far seen two calculations of anomalies. One involving a one-loop calculation
in two dimensions, and the other the calculation of Fujikawa which involves regularizing
the determinant from a change of variables in the fermionic path integral in four dimensions.
This is essentially a one-loop calculation as well given the well-known diagrammatic interpration
of fermion determinants. There are many qualitative reasons why we would expect that
these one-loop calculations give the anomaly exactly, without any corrections from higher
orders in perturbation theory. As will see in later sections, there are several topological
interpretations of the anomaly, one involving the topology of gauge configuration space and
another involving the index of the Dirac operator. This argues against perturbative corrections
since topological quantities cannot change continuously.

The detailed diagrammatic proof of the the absence of higher-order corrections to the
anomaly was given by Adler and Bardeen \cite{AdlerER}. The essential idea in their proof was
to note that higher-order corrections necessarily involve internal boson propagators, and then
to show that these can always be regularized in such a way that the usual Ward identities
are satisfied.

\subsection{Why anomalies are an infrared effect}

The calculations we have given of the anomaly so far seem to focus on the issue
of regularization. The anomaly reflects the fact that the UV regulator does not respect the chiral
symmetry of the theory, hence the lack of conservation of the chiral current. This suggests
that the anomaly is an ultraviolet effect. However, it later came to be appreciated that anomalies
are more accurately understood as a statement about the infrared behavior of the theory
\cite{tHooft,fsby,ColemanGrossman}.  First of all, only massless particles for which no mass term
is allowed that is consistent with the potentially anomalous symmetry can contribute to the anomaly. If a mass term
were allowed, then the theory could be regulated in a way that respects the symmetry using the Pauli-Villars
method, and so would not be anomalous.

More importantly, the anomaly can be understood as a statement about the analytic structure
of current correlation functions. The anomaly equation implies the existence of discontinuities
in current correlation functions at zero momentum. Only massless particles can contribute to
these discontinuities, so the anomaly of a theory can always be understood purely in terms of
the spectra and interactions of the massless states. We saw a simple example of this in our
analysis of the chiral anomaly in $1+1$ dimensions. There, the value of the coefficient $B$ in
\calle{onenn} was determined by the massless fields and this non-zero value, along with
gauge invariance, required an anomaly in the divergence of the axial current. For more details on
this point of view see \cite{tHooft,fsby,ColemanGrossman}.

\subsection{Exercises for Lecture 1}

\begin{itemize}
\item In $1+1$ dimensions a $U(1)$ chiral current  $j(z)$ obeys the Operator Product
Expansion
\beq
\label{proba}
j(z) j(0) \sim  {k \over z^2} 
\eeq
Compute the anomalous divergence of the current when coupled to a background gauge field
in terms of the constant $k$. For help see sec. 12.2 of \cite{Poltwo}.
\item Compute the vacuum polarization diagram in $1+1$ dimensions using a momentum-space
cutoff and show that the coefficient $A$ in \calle{onemm} is logarithmically divergent and that
the value you get for the coefficient $B$ in \calle{onemm} is the same as found using dimensional
regularization.
\end{itemize}

\section{Lecture 2: Applications of anomalies in $D=4$}

The results of the previous lecture on the evaluation of the chiral anomaly in four dimensions  can be generalized in several different ways. In this lecture we consider these generalizations and some
of their ``real-world" applications to four-dimensional physics.

First of all, we can replace electromagnetism by a general gauge theory in the previous analysis.  Redoing the Fujikawa analysis for fermions coupled to a non-Abelian gauge theory leads to
a simple change in the final answer:  there is a trace over gauge indices so that $F_{\mu \nu} F_{\alpha \beta}$ is
replaced by $2\Tr (F_{\mu \nu} F_{\alpha \beta})$ assuming the standard normalization $ \Tr \lambda^a \lambda^b = {1 \over 2} \delta^{ab}$ for the generators of the gauge symmetry.

We can also generalize the Abelian chiral current to a set of  non-Abelian chiral currents 
\beq
\label{zerocc}
j_\mu^{Ai}= \bar \psi \gamma^\mu \gbar T^i \psi
\eeq
with $T^i$ the generators of the chiral symmetry. The divergence of $j_\mu^{Ai}$ then
involves an additional trace over the generator $T^i$. Combining these two results, the
anomalous divergence of a non-Abelian axial current coupled to non-Abelian gauge fields
is
\beq
\label{zerocd}
\partial^\mu j_\mu^{Ai} = - {1  \over 16 \pi^2} \epsilon^{\alpha \beta \mu \nu} \Tr ~T^i F_{\alpha \beta}F_{\mu \nu}.
\eeq

\subsection{$\pi^0 \rightarrow \gamma \gamma$ revisited}

As an example of these generalizations of  phenomenological importance we will revisit the
problem of $\pi^0$ decay \cite{Adler,Bell} discussed in sec. 1.1 and redo the analysis in QCD. Thus, we consider
QCD with two flavors in the limit $m_u=m_d=0$ and write the quark doublet as
\beq
\label{zerodd}
Q=\pmatrix{u \cr d}.
\eeq
This theory has an $SU(2)_L \times SU(2)_R$ chiral symmetry with a triplet of axial currents
given by
\beq
\label{zeroee}
j_\mu^{A i} = \bar Q \gamma_\mu \gbar \tau^i Q
\eeq
with $\tau^i$ the generators of $SU(2)$. The divergence of $j_\mu^{Ai}$ has a contribution
from external gluons which is proportional to $\Tr \tau^i \lambda^c \lambda^d =0$ with $\lambda^a$ the generators of
the $SU(3)$ gauge symmetry of QCD. This vanishes since the $\tau^i$ are traceless, hence there is no QCD contribution to the anomaly. On the
other hand, the electromagnetic contribution is given by
\beq
\label{zeeroff}
\partial^\mu j_\mu^{A i} |_{QED} = - {1 \over 16 \pi^2} \epsilon^{\alpha \beta \mu \nu}
F_{\alpha \beta} F_{\mu \nu} \Tr (\tau^i Q_{el}^2 )
\eeq
where
\beq
\label{zerogg}
Q_{el} = \pmatrix{ 2/3 & 0 \cr 0 & -1/3 } 
\eeq
is the electric charge matrix acting on the quark doublet $Q$. This  trace is non-zero only
for $i=3$, and this component of the axial current has a divergence 
\beq
\label{zerohh}
\partial^\mu j_\mu^{A 3} |_{QED} = - {1 \over 32 \pi^2} \epsilon^{\alpha \beta \mu \nu}
F_{\alpha \beta} F_{\mu \nu}. 
\eeq
Using the fact that $j_\mu^{A3}$ creates a $\pi^0$, this anomaly implies an effective
coupling of the $\pi^0$ to electromagnetism of the form \calle{zerob} with coefficient
\beq
\label{zeroii}
A = {e^2 \over 16 \pi^2 f_{\pi}}.
\eeq
where $f_\pi = 93 ~{\rm MeV}$ is the pion decay constant. 
This gives a lifetime for $\pi^0 \rightarrow \gamma \gamma$ in excellent agreement with
experiment.

\subsection{Cancellation of gauge anomalies}

So far we have considered situations where  currents of global axial symmetries are anomalous
in the presence of gauge fields coupled to vectorial currents. In
the standard model and its extensions to grand unified theories
and string theory we are interested in situations where the gauge
currents are themselves chiral. We will not yet delve into the details of such gauge anomalies,
putting this off for later when we have developed more of the necessary formalism. However,
we do have the tools to see when such gauge anomalies vanish.

We focus here on the situation in $D=4$. In this case we can always write
the Lagrangian purely in terms of left-handed fields since given
a right-handed fermion field $\psi_{R, \alpha}$, the field $\tilde \psi_{L,\alpha}
= \epsilon_{\alpha \beta} \psi_{R \beta}^*$ transforms as a left-handed
fermion under Lorentz transformations. The action for fermion fields  in
the representation ${\bf r}$ of the gauge group
has the form
\beq
\label{onezz}
{\cal L}_f = \psi_{L i}^\dagger i \bar \sigma_\mu D^\mu_{ij}
\psi_{Lj} + \cdots
\eeq
where spinors indices have been suppressed, $i= 1 \cdots {\rm dim} ~{\bf r}$
and the covariant derivative is given by
\beq
\label{zy}
D^\mu_{ij} = \partial^\mu \delta_{ij} - i g A^{\mu a }(\lambda_{\bf r}^a)_{ij}
\eeq
with $\lambda_{\bf r}^a$ the generators of the Lie algebra of the gauge group.

We then need to compute the divergence of the gauge current, which is given
by a triangle diagram with the current at one vertex and gauge fields on the other
two vertices.  Label the  vertex with the current by $\mu, a$.  Bose symmetry  demands that the diagram
is invariant under the exchange of the two external gauge fields. The divergence
of the current is thus proportional to $\Tr (\lambda_{\bf r}^a \{ \lambda_{\bf r}^b, \lambda_{\bf r}^c \} )$
and vanishes if and only if
\beq
\label{zx}
 d^{abc} \equiv \Tr (\lambda_{\bf r}^a \{ \lambda_{\bf r}^b,\lambda_{\bf r}^c \}) = 0.
 \eeq
This quantity vanishes if the fermion representation ${\bf r}$ is real (or pseudo-real). There
are several ways to understand this. Physically, if the fermions are in a real representation then it is possible to add a gauge invariant mass term to the Lagrangian. We can thus regulate the theory in a gauge invariant way using Pauli-Villars regularization and hence the anomaly must vanish. We can also show this directly. 

A field $\psi_{\bf r}$ in the representation $\br$ transforms under infinitesimal gauge transformations by
\beq
\label{oneb}
\psi_\br \rightarrow (1+i \alpha^a \lambda_\br^a) \psi_\br.
\eeq
Therefore the complex conjugate field transforms as
\beq
\label{onec}
(\psi_\br)^*\rightarrow (1-i \alpha^a ( \lambda_\br^a)^* )( \psi_\br)^*
\eeq
which shows that the matrices which represent the conjugate representation $\bar \br$ are
$\lambda_{\bar \br}^a = -(\lambda_\br^a)^* = -(\lambda_\br^a)^T$ where in the last step we have used the fact that we
can choose the $\lambda_\br^a$ to be Hermitian. 

Now if the representation $\br$ is real or pseudo-real then we can find a
unitary matrix $S$ such that 
\beq
\label{oned}
\lambda_\br^a = S \lambda_{\bar \br}^a S^{-1} = S (-(\lambda_\br^a)^T) S^{-1}.
\eeq
In this case we have
\beq
\label{oneee}
d^{abc}=\Tr \lambda_\br^a \{\lambda_\br^b,\lambda_\br^c\} = -\Tr (\lambda_\br^a)^T \{(\lambda_\br^b)^T,(\lambda_\br^c)^T\} =-\Tr \{\lambda_\br^b,\lambda_\br^c\} \lambda_\br^a =-d^{abc}
\eeq
from which we conclude that $d^{abc}=0$. 

Groups which have only real or pseudo-real representations and hence no gauge anomalies
are $SU(2)$, $SO(2n+1)$ for $n \ge 2$, $Sp(2n)$ for $n \ge 3$ and $G_2$, $F_4$, $E_7$ and $E_8$.
The remaining Lie groups in Cartan's classification, $U(1)$,$SU(n)$ for $n \ge 3$, $E_6$ and
$SO(4n+2)$ all have potential anomalies. Except for $U(1)$, the groups with potential anomalies are also those groups for which $\pi_5(G) \ne 0$, a fact which is related to a topological characterization of anomalies which is reviewed in \cite{agg} and described briefly in sec 3.6. 

\subsection{'t Hooft matching conditions}

One of the more useful applications of anomalies in $D=4$ arose in a study of the bound state
spectrum of confining gauge theories by 't Hooft \cite{tHooft}. 
Given that nuclei are bound states of neutrons
and protons and neutrons and protons in turn bound states of quarks, it is natural to ask whether
quarks, or quarks and leptons, could themselves be bound states of some other objects (often called
preons). On obvious objection to this idea is that there is no experimental evidence for substructure
up to around the TeV scale, while the masses of light quarks and leptons are in the MeV range,
much lighter than any possible scale of compositeness. How could there possibly be such light
bound states if the scale of confinement or binding is so large? Such a theory would have to behave
much differently from QCD where the mass scale of 
hadrons is comparable to the scale $\Lambda_{\rm QCD}$ where QCD becomes strongly interacting.

The only plausible mechanism that could give rise to such light states would be a theory which confines but does not break chiral symmetries.
The bound state fermions could then be light because of the unbroken chiral symmetries. The argument
of 't Hooft, based on anomalies, strongly constrains such a possibility. His argument goes as follows:

Consider a gauge theory with chiral fermions and an unbroken, anomaly-free, global symmetry group ${\cal G}$. Anomaly-free
means that there are no ${\cal G}$---gauge---gauge anomalies so that ${\cal G}$ is a valid
symmetry even in the presence of background gauge fields. Suppose further that the triangle diagram
with three ${\cal G}$ currents is anomalous, that is the the coefficients $d^{ijk}$ are non-zero, where $i,j,k$ run over $1 \ldots {\rm dim} ~{\cal G}$.

Now imagine that we add a set of massless, gauge-singlet (spectator) fermions which contribute $ - d^{ijk}$
to the ${\cal G}^3$ anomaly. Once we have done this we could gauge ${\cal G}$ because it is now
completely free of anomalies. Imagine we have done this.  Now, if the orginal gauge theory confines,
we could study the low-energy effective action that describes the massless excitations of the resulting
theory.
In general this theory could contain massless Nambu-Goldstone bosons, but since we assume
that ${\cal G}$ is not spontaneously broken, there are no such bosons that are relevant to our analysis
of ${\cal G}$.  The spectator fermions will remain massless since they were massless originally
and being gauge singlets, are not affected by the dynamics of confinement. And then finally, there
could be massless bound states.

Now the effective low-energy theory must be consistent since we started with a consistent
theory. But the only way it can be consistent, that is anomaly free, is if the anomaly of the
spectator fermions is canceled by an anomaly coming from massless bound states.
Thus we are led to 't Hooft's conclusion, that there must be a set of massless bound states which 
have the same anomaly $d^{ijk}$ as the original fundamental fields. 

This argument did not involve the value of the gauge coupling when we gauged ${\cal G}$, so we
could just as well take it to be zero. This decouples the ${\cal G}$ gauge fields,  so we conclude that 't Hooft's condition must also be true in the original theory with
${\cal G}$ a global symmetry and no spectator fermions.
For a version of this argument which relies more heavily on the analytical aspects of the anomaly
see \cite{fsby}. 

I will give one application of this condition. Consider QCD with three flavors of quarks
in the massless limit $m_u=m_d=m_s=0$. This theory has a ${\cal G}=SU(3)_L \times SU(3)_R 
\times U(1)_V$ global symmetry which has no anomalies involving gauge currents. There
are however $SU(3)_R$---$SU(3)_R$---$U(1)_V$ and $SU(3)_L$---$SU(3)_L$---$U(1)_V$ 
anomalies
as well as $SU(3)_L^3$ and $SU(3)_R^3$ anomalies.
Assume that QCD confines into color singlets without spontaneous breaking of ${\cal G}$.
Then you can check that there is no massless bound state spectrum which satisfies the
't Hooft anomaly matching conditions. Therefore the assumption that QCD confines without
breaking ${\cal G}$ must be false.  Of course we know in the real world that chiral
symmetry is spontaneously broken. What is interesting about this argument is that it shows
that in part this is unavoidable. It isn't simply a consequence of complicated dynamics which
could have happened one way or the other. Rather, it follows from consistency of the theory.
To be clear, anomalies in this model do not completely dictate the pattern of chiral
symmetry breaking observed in the real world, they simply say that not all of the chiral
symmetry can remain unbroken.

\subsection{Exercises for Lecture 2}

\begin{itemize}
\item  Verify that all anomalies cancel in the Standard Model
and also for $G=SU(5)$ with fermions in the representation
$\bar 5 \oplus 10$ (the latter fact of course implies the former).
\item  't Hooft's matching conditions play an important role in checking the consistency of
certain dual descriptions of gauge theories.  Work through the arguments in \cite{SeibergPQ}
to determine the anomaly free (that is no gauge anomaly) global chiral symmetries and check that the
cubic global anomalies match for the proposed Seiberg duals of supersymmetric 
generalizations of QCD. 
\end{itemize}

\section{Lecture 3: Mathematical aspects of anomalies}

One of the themes running through these lectures will be a relation
between gauge anomalies in $2n$ dimensions and chiral anomalies in
$2n+2$ dimensions. By gauge anomalies we mean the anomaly in the divergence of a current coupled to a gauge field alluded to in sec. 2.2. but not worked out in detail. By the chiral anomaly we mean the
anomaly in the divergence of the axial current. In this section we will briefly explore the
mathematical origin of this connection. In the fourth lecture we
will discuss a physical model which provides a more direct interpretation
of the result.

The objects such as $\Tr(F^n)$ which we have encountered in the
study of anomalies are examples of a general class of mathematical
objects called characteristic classes. Manipulations with characteristic
classes play an important role in the study of anomalies and in the relation between
gauge and chiral anomalies, so we will
take a detour here to review some of the basic material used in
the study of these objects. Further details may be found in
\cite{egh}.

\subsection{Characteristic classes}

To start, suppose that $\alpha$ is a $k \times k$ complex matrix and
$P(\alpha)$ is a polynomial in the components of $\alpha$. We can
act on $\alpha$ by elements $g$ of $GL(k,\IC)$,
\beq
\label{zw}
g: \alpha \rightarrow \alpha^g = g^{-1} \alpha g.
\eeq
We will say that $P(\alpha)$ is a characteristic polynomial if
$P$ is invariant under $GL(k,\IC)$ transformations on $\alpha$,
that is if $P(\alpha^g) = P(\alpha)$. We can also consider
characteristic polynomials for subgroups of $GL(k,\IC)$ such
as $U(k)$, $GL(k, \IR)$, $O(k)$ and $SO(k)$.

Examples of characteristic polynomials are easy to come by. The canonical
example arises by expanding the determinant
\beq
\label{zv}
\Det(1 + \alpha) = 1 + S_1(\lambda) + S_2(\lambda) + \cdots + S_k(\lambda).
\eeq
where $\lambda_1, \cdots \lambda_k$ are the eigenvalues of $\alpha$
and $S_j(\lambda)$ is the  $j^{th}$ symmetric polynomial,
\beq
\label{zu}
S_j(\lambda) = \sum_{i_1<i_2< \cdots i_j} \lambda_{i_1}
\lambda_{i_2} \cdots \lambda_{i_j} .
\eeq
In general characteristic polynomials for $GL(k,\IC)$ are
expressible as polynomials in the $S_j(\lambda)$.

To obtain characteristic classes we substitute $k \times k$
matrix valued $2$-form curvatures for $\alpha$. We will denote
a generic curvature and connection by $(\Omega,\omega)$. When
we have specific examples to discuss we will use the gauge
curvature and connection $(F,A)$  or the gravitational curvature
and connection $(R, \omega)$. 

We first consider complex bundles with $g \in GL(k,\IC)$ or
$g \in U(k)$. The characteristic polynomials are the same, and from
the physics point of view $U(k)$ is more natural since we naturally
use compact gauge groups in physics. The curvature $\Omega$ can be taken to be
anti-Hermitian. Substituting $i \Omega / 2 \pi$ for $\alpha$ in \calle{zv}
gives the total Chern form
\beq
\label{zt}
c(\Omega) = \Det(1 + {i \Omega \over 2 \pi}) =
1 + c_1(\Omega) + c_2(\Omega) + \cdots + c_k(\Omega)
\eeq
where the $j^{th}$ Chern form is a polynomial in $\Omega$
of degree j:
\beqn
\label{zs}
c_1(\Omega) & = & {i \over 2 \pi} \Tr \Omega \\
c_2(\Omega) & =  & {1 \over 8 \pi^2} (\Tr \Omega^2 -
                 (\Tr \Omega)^2 )
\eeqn
and so on.
If we consider instead real bundles with transition functions in
$GL(k,\IR)$ or $O(k)$ then the curvature $\Omega$ is a real antisymmetric
matrix and it is natural to consider the total Pontrjagin class
defined by
\beq
\label{zr}
p(\Omega) = \Det(1 - {\Omega \over 2 \pi}) = 1 + p_1(\Omega)
+ p_2(\Omega) + \cdots+p_k(\Omega).
\eeq

In physics applications we will typically encounter Chern classes in
computing topological invariants of $U(k)$ gauge theory and Pontrjagin
classes  when we compute topological invariants of gravitational theories
where $O(k)$ or $O(k-1,1)$, depending on signature, act as local Lorentz
transformations.

For real bundles $\Omega$ is an antisymmetric $k \times k$ matrix.
If $k$ is an even integer, $k=2r$, then we can put $\Omega$ in
the form
\beq
\label{zq}
{i \Omega \over 2 \pi} = \pmatrix{ 0 & x_1 &  & & & \cr
                           -x_1 & 0 &  & & & \cr
                              &  & . & & & \cr
                              &  &  & . & &  \cr
                            & & &  & 0 & x_r\cr
                            & & & & -x_r & 0 \cr }   
\eeq
in which case we have
\beqn
\label{zp}
p_1 & = & \sum_a x_a^2 \\
                 p_2 & =  & \sum_{a<b} x_a^2 x_b^2 
 \eeqn
and so on. If $k$ is an odd integer then we can put $\Omega$ in the form \calle{zq} but
with an extra row and column of zeroes. 

With $k=2r$ if we demand only $SO(2r)$ invariance rather than
$O(2r)$ invariance (that is we have an oriented real vector bundle)
then there is one additional characteristic class we can define.
If $\alpha$ is a $2r \times 2r$ antisymmetric matrix, then putting
$\alpha$ in the form \calle{zq} we can define the Pfaffian of $\alpha$
by
\beq
\label{zo}
Pf(\alpha) = x_1 x_2 \cdots x_n.
\eeq
Substituting the curvature $\Omega$ for $\alpha$ gives a characteristic
class called the Euler class $e(\Omega)$. Note that $e$ is the square
root of the top Pontrjagin class, $e^2(\Omega) = p_k(\Omega)$.

Combinations of characteristic classes often appear in the computation of
topological invariants. These include the A-roof genus
given by
\beq
\label{zm}
\hat A(R) = \prod_a {x_a/2 \over \sinh{x_a/2} } =
1 - {p_1 \over 24} + 
{1 \over 16}\left( {7 \over 360}p_1^2 - {1 \over 90} p_2 \right)+ \cdots 
\eeq
and the Hirzebruch L-polynomial
\beq
\label{zj}
 L(\Omega) = \prod_j {x_j \over \tanh{x_j} }  = 1 + {1 \over 3} p_1 + {1 \over 45} \left( 7 p_2 - p_1^2 \right) + \cdots.
 \eeq

For later use we also mention the behavior of characteristic classes
for sums of vector bundles. Given two vector bundles $E$, $F$
with connections there is a natural vector bundle $E \oplus F$ called
the Whitney sum with the curvature of $E \oplus F$ the direct sum of
the curvatures of $E$ and $F$. The total Chern and Pontrajin classes
obey the relations
\beq
\label{zjj}
P(E \oplus F) =p(E) p(F), \qquad c(E \oplus F) = c(E) c(F) 
\eeq
thus
\beq
\label{zjjj}
p_1(E \oplus F) = p_1(E) + p_1(F), \qquad
          p_2(E \oplus F) = p_2(E) + p_2(F) + p_1(E) p_1(F), \quad
 \eeq
and so on.
\subsection{Properties of characteristic classes}

Having introduced characteristic classes, we now need to study some of their properties.
In particular we would like to show
\begin{itemize}
\item That $P(\Omega)$ is closed.
\item  That  integrals of $P(\Omega)$ are topological invariants.
\end{itemize}


To show that $P$ is closed it suffices to show this for
$P_m \equiv  \Tr \Omega^m $  since a general invariant polynomial
can be expressed in terms of sums and products of the $P_m$. In this section
we denote a generic connection by $\omega$ and its curvature by $\Omega$. 
Using the chain rule we have
\beq
\label{zh}
d P_m = m \Tr d \Omega \Omega^{m-1} 
\eeq
which we can write using the Bianchi identity,
$D \Omega = d \Omega + \omega \Omega - \Omega \omega =0$, as
\beq
\label{zg}
d P_m = m \Tr (\Omega \omega - \omega \Omega) \Omega^{m-1} =0
\eeq
where in the last step we have used cyclicity of the trace.

To show that the integrals of $P$ are topological invariants it will
again be sufficient to establish this for $P_m$. To do this we consider
two connections $\omega_0,\omega_1$ with the same transition functions
and will show that the difference $P_m(\Omega_1)-P_m(\Omega_0)$ is exact.
The integral of $P_m$ over a closed manifold is then independent of
variations of the connection (while keeping the transition functions
fixed) and is thus a topological invariant.

It is useful to construct an interpolation between the two connections.
Let $t \in [0,1]$ be a real parameter and consider the connection and
curvature
\beqn
\label{zf}
\omega_t & =  & \omega_0 + t (\omega_1 - \omega_0 ) , \\
                 \Omega_t & = & d \omega_t + \omega_t^2 
\eeqn
A small amount of algebra shows that
\beq
\label{ze}
{\partial \Omega_t \over \partial t} = d(\omega_1 - \omega_0)
+ [ \omega_t, \omega_1 - \omega_0 ] = D_t (\omega_1 - \omega_0) ,
\eeq
with $D_t$ the covariant derivative with respect to $\omega_t$.
We thus have
\beq
\label{zd}
{\partial \over \partial t} P_m(t) = m \Tr {\partial \Omega_t
\over \partial t} \Omega_t^{m-1} = m \Tr D_t (\omega_1 - \omega_0)
\Omega_t^{m-1} = m d \Tr (\omega_1 - \omega_0) \Omega_t^{m-1} 
\eeq
where we have used the Bianchi identity to pull $D_t$ out of the trace
and then gauge invariance to reduce $D_t$ to $d$ acting on the trace.
Integrating \calle{zd} with respect to $t$ then yields the desired result
\beq
\label{zc}
P_m(\Omega_1) - P_m(\Omega_0) = m d
\int_0^1 dt \Tr (\omega_1 - \omega_0) \Omega_t^{m-1} .
\eeq

In applications to anomalies the most important fact about characteristic classes
is that the obey a set of equations called the descent equations.  I will prove these
equations for 
characteristic classes constructed from the connection and curvature in
gauge theory. It turns out that a natural and elegant way to do this involves
the use of ghosts and the BRST formalism. To explain why the BRST formalism
is particulary useful in the study of anomalies I first take a short detour
by introducing the Wess-Zumino consistency conditions and then  translating
these into the language of BRST cohomology. 

\subsection{WZ condition and BRST cohomology}

We have found that the path integral over fermion fields can lead to a violation of
gauge invariance in theories with chiral gauge currents. Following Wess
and Zumino\cite{wz} we will first show that this violation obeys a certain
condition. The effective action that results from integrating out the
fermions is defined by
\beq
\label{zb}
e^{-W[A]} = \int D \psi D \bar \psi e^{- \int \bar \psi i \Dslash \psi}.
\eeq
Under an infinitesimal gauge transformation we have 
\beq
\label{za}
\delta_v A = dv + \[A,v \] = Dv,
\eeq
so the gauge variation of $W[A]$ can be written as
\beqn
\label{ya}
\delta_v W[A] & = & W[A+Dv]-W[A] = 
\int (D_\mu v)^a {\delta W[A] \over \delta A_\mu^a} \\
                                 &= & - \int v^a \left( D_\mu 
                                 {\delta W[A] \over \delta A_\mu}\right)^a
                                 =-\int v^a D^\mu J_\mu^a \\
                                 &= & - \int v^a {\cal A}^a[x,A] 
\eeqn
where the current is $J_\mu^a = \delta W/\delta A^{\mu a}$ and since we choose gauge
variations which vanish at infinity, we have also freely integrated by parts.

The generator of gauge transformations acting on functionals of the gauge field
is thus
\beq
\label{yb}
v^a D_\mu {\delta \over \delta A_\mu^a (x)} \equiv - v^a \CJ^a(x).
\eeq
A short calculation shows that the generators obey the algebra
\beq
\label{yc}
\[\CJ^a(x),\CJ^b(y) \] = i f^{abc} \delta^4(x-y) \CJ^c(x) 
\eeq
and this, along with the definition of the anomaly $\CA^a$ as the covariant
divergence of the current implies the Wess-Zumino condition
\beq
\label{yd}
\CJ^a(x) \CA^b[y,A]-\CJ^b(y) \CA^a[x,A]= i f^{abc} \delta^4(x-y) \CA^c[y,A] .
\eeq

This condition obviously follows  just from the algebra of gauge variations and the fact
that the anomaly arises from variation of an effective action, $W[A]$, but nonetheless it is quite useful.
Historically it was useful in sorting out the proper calculation of gauge anomalies
from Feynman diagrams. More recently it has played an important role in understanding
the cohomological interpretation of anomalies as we now discuss.

We will introduce ghost fields and the BRST operator into our discussion. There are
at least three reasons for doing this. First, to actually make sense of the path integral
in gauge theories we need to gauge fix and replace gauge invariance by BRST invariance. 
Thus we should really formulate gauge anomalies in this language as well. Second, it
is technically useful in demonstrating the descent equations, and third there is an
elegant interpretation of the Wess-Zumino consistency condition in this language \cite{Baulieu,Brandt}.

Recall that in the BRST formalism we introduce ghosts $\omega^a(x)$ which are Grassmann
valued spin zero fields in the adjoint representation of the gauge group (there are
also anti-ghosts but they will not play in the following discussion). The BRST operator
${\CS}$ acts on the gauge fields and ghosts as
\beqn
\label{ye}
{\CS} A_\mu^a & = & \partial_\mu \omega^a + f^{abc} A_\mu^b \omega^c, \\
                 {\CS} \omega^a & = &  -{1 \over 2} f^{abc} \omega^b \omega^c .
\eeqn
Note that the action on $A_\mu^a$ is that of a gauge transformation, but with the ghost
field as
anti-commuting gauge parameter. Geometrically ghosts should be thought of as $1$-forms
in the space $\CG$ of gauge transformations \cite{bcr}. Consistent with this we will write
the gauge field as a Lie-algebra valued $1$-form $A = A_\mu^a \lambda^a dx^\mu$ and use a formalism
where the $1$-forms on spacetime, $dx^\mu$, anticommute with the $1$-forms $\omega^a$
on $\CG$:
\beq
\label{yf}
dx^\mu \omega^a(x) + \omega^a(x) dx^\mu = 0, 
\eeq
and similarly we will take the BRST operator ${\CS}$ and the exterior derivative
$d$ to be anticommuting:
\beq
\label{yg}
{\CS} d + d {\CS} = 0.
\eeq
In fact, ${\CS}$ can be thought of as the exterior derivative on $\CG$. The 
gauge transformation law of the ghost field can then be interpreted as the
Mauer-Cartan equation on $\CG$. 

As a first application of this formalism we consider the anomaly with the anomaly
parameter $v^a$ replaced by the ghost field $\omega^a$,
\beq
\label{yh}
\CA[\omega,A] = \int d^4x ~ \omega^a(x) \CA^a[x,A] .
\eeq
A short calculation shows that the Wess-Zumino consistency condition
\calle{yd} is equivalent to the statement that ${\CS} \CA[\omega,A]=0$, that is that
the anomaly \calle{yh} is BRST closed. Since ${\CS}^2=0$, this can obviously be
satisfied by setting $\CA = {\CS} F[A]$, but then we could simply add $F[A]$
to the action and cancel the anomaly. Since anomalies are defined precisely up to
the choice of regulator, or equivalently up to the ability to add such a local counterterm 
to the action (see e.g. the discussion
in sec 1.2), possible anomaly terms
are classified by the cohomology of the BRST operator (in the space of local
functionals) at ghost number one.

\subsection{Descent formalism for anomalies}

We now have the necessary machinery to demonstrate the descent equations.
These state that if 
$\alpha_{2n+2}$ is a characteristic $2n+2$-form then we have locally 
\beqn
\label{zi}
\alpha_{2n+2} & = & d \alpha^{(0)}_{2n+1} \\
                 \delta \alpha^{(0)}_{2n+1} & = & d \alpha^{(1)}_{2n} \\
                 \delta \alpha^{(1)}_{2n} & = & d \alpha^{(2)}_{2n-1}
\eeqn
In these equations $\delta$ indicates the gauge variation, and the superscript indicates
whether the quantity is independent of the parameter of the gauge variation (0), first
order in the gauge parameter (1), and so on.

The descent formula follows quite easily from the BRST formalism and we will see as an
extra bonus that the BRST formalism also gives us a natural candidate for the gauge anomaly.
We will prove the descent formalism
for the particular case that $P(F)= \Tr F^{n+1} \equiv \alpha_{2n+2}$. Other characteristic
classes for the gauge field can be written as sums and products of these so
it suffices to show the descent formalism for this case.

First, we have already seen that $\alpha_{2n+2}$ is closed. Locally, or on a simply
connected manifold, this implies the first step in the descent formula,
\beq
\label{yi}
\alpha_{2n+2} = d \alpha_{2n+1}^{(0)} . 
\eeq
Now since $\alpha_{2n+2}$ is gauge invariant and independent of ghost fields, we
also have ${\CS} \alpha_{2n+2}=0$. Turning to the next step in the descent formalism,
we consider the gauge variation of $\alpha_{2n+1}^{(0)}$ with an anti-commuting gauge
parameter, that is ${\CS} \alpha_{2n+1}^{(0)}$. This is also closed because
\beq
\label{yj}
d {\CS} \alpha_{2n+1}^{(0)} = -{\CS} d \alpha_{2n+1}^{(0)}=
-{\CS} \alpha_{2n+2} = 0. 
\eeq
Therefore we have shown that, locally, 
\beq
\label{yk}
 \CS \alpha_{2n+1}^{(0)} = d \alpha_{2n}^{(1)} 
 \eeq
where $\alpha_{2n}^{(1)}$ has ghost number one. We can continue this process, for example
we also have
\beq
\label{yl}
d {\CS} \alpha_{2n}^{(1)} = -{\CS} d \alpha_{2n}^{(1)} = 
-{\CS}^2 \alpha_{2n+1}^{(0)} = 0 ,
\eeq
which shows that
\beq
\label{ym}
 {\CS} \alpha_{2n}^{(1)} = d \alpha_{2n-1}^{(2)} .
 \eeq

Equations \calle{yi}, \calle{yk}, and \calle{ym} constitute a demonstration that $\alpha_{2n+2}$ obeys the descent equations. We have expressed these equations in the BRST formalism, but having derived them,
we can replace the BRST variation by the gauge variation and the ghost fields by parameters
of the gauge variation and they still hold true.  In addition, \calle{ym} shows that
the spacetime integral of $\alpha_{2n}^{(1)}$ provides a candidate for
the anomaly in $2n$ spacetime dimensions since it is BRST closed and has ghost number
one. In fact, the spacetime integral of $\alpha_{2n}^{(1)}$
is precisely the gauge anomaly in $2n$ dimensions up to numerical factors which will
be discussed in the following lecture where a physical model of this connection will be 
presented. 
Note that if we work strictly  in $2n$ dimensions we would have to add two extra dimensions to
carry out this procedure in order to make sense of $2n+2$-forms.

\subsection{Determinant line bundle}

Anomalies have a deep connection to the topology of the configuration space of
gauge theories. This connection was understood in \cite{as,zum,stora,AlvarezGaumeCS}.
A very useful review
can be found in \cite{agg} and as a result I will be rather brief here.

We are interested in anomalies in the effective action which arises
from integrating out the fermions in a theory coupled to gauge theory
or gravity. We consider left-handed Weyl fermions in some
representation ${\bf r}$ of a gauge group $G$. In $D= 0 \ {\rm mod} \ 4$
dimensions we write all fermions fields as left-handed since
complex conjugation relates left- and right-handed
fields. In $D=2 \ {\rm mod} \ 4$ dimensions this is not the case and
right-handed fermions must be treated separately from left-handed
fermions. The results will however only differ from the analysis
below by the overall sign of the anomaly. 
The fermion effective action is
\beq
\label{twa}
e^{-W[A]} = \int {\cal D} \psi {\cal D} \bar \psi
e^{- \int d^{2n} x \bar \psi i \Dslash_+ \psi}    
\eeq
One often writes $e^{-W[A]} = \det i \Dslash_+$ but this is not
completely correct. To define a determinant one needs an operator
which maps a vector space to itself while $\Dslash_+$ maps left-handed
fermions to right-handed fermions. This can be dealt with in various
ways. One possibility discussed in \cite{agg} is to add to $\Dslash_+$
a free operator $\pslash_-$ which acts on right-handed fermions
and to define the effective action to be the determinant of
$\Dslash_+ + \pslash_-$. This changes the effective action only
by an overall constant which is independent of the gauge fields
and so does not affect any possible anomalies in the gauge
variation of the efective action.

Another subtlety can appear in $D=2,10$ where one can have real
Majorana-Weyl fermions. In this case integrating out the fermions
give the square root of the determinant or more precisely the
Pfaffian of $\Dslash_+ + \pslash_-$. This will result in factors
of $1/2$ is the formula for the anomaly.

If we consider fermions in the representation ${\bf r}$
and in the complex conjugate representation ${\bf \bar r}$ then
it follows from the form of the action that $W_{\bar \br}[A] =
(W_{\br}[A])^*$. Thus
\beq
\label{twd}
2 {\rm Re} W_{\br}[A] = W_{\br}[A] + W_{\bar \br}[A] =
W_{\br \oplus \bar \br}[A] .
\eeq
Since $W_{\br \oplus \bar \br}[A]$ is the effective action for a fermion
in a real representation, it can be regulated in a gauge invariant
way. Therefore there can be no anomaly in the real part of the
effective action so the only anomaly can be in the phase of
$\det i \Dslash$.

In gauge theory we have the space of all gauge connections ${\cal A}$, and
the group of gauge transformations ${\cal G}$ (these are maps from spacetime into
the gauge group $G$). 
To define
the fermion effective action  we must define $e^{-i {\rm Im} W[A]}$
for each point in ${\cal C}= {\cal A} /{\cal G}$, since as discussed above, the real part
of $W[A]$ is clearly well defined.  Now $e^{-i {\rm Im} W[A]}$ is a
phase, that is an element of the group $U(1)$. We should view this
in the language of principal fibre bundles. The base space is ${\cal C}$
and we try to patch together $U(1)$ fibres to define the effective
action. Now if we can assign a unique value to the effective action
for each point in ${\cal C}$ then this $U(1)$ bundle has a global
section. Conversely, if there is no global section then we cannot
uniquely define the effective action for each point in the configuration
space and the theory is anomalous.

Now the existence of a global section is equivalent to the statement
that the $U(1)$ bundle is trivial. Thus we have turned the question of
anomalies into a topological question since obstructions to the
triviality of the bundle are topological in nature.

A famous example of a topologically non-trivial $U(1)$ bundle is the
Dirac-Wu-Yang monopole bundle over $S^2$. This suggests that we
look for a non-trivial two-sphere in ${\cal C}$. It should be
noted that ${\cal A}$ is contractible since any gauge field
can be shrunk down to zero continuously. Thus ${\cal A}$ has trivial
topology and any non-trivial topology of ${\cal C}$ must arise
from taking the quotient by ${\cal G}$. An analysis along these lines
relates the existence of a global section to the question of whether or
not $\pi_5(G)$ is non-trivial or not (for theories in four dimensions) and eventually leads to an identification between
the chiral anomaly and the gauge anomaly as given by the descent procedure, a connection which 
will be explained in the following lecture. 
For further details on this approach
see \cite{agg}. 

\subsection{The Dirac index and the chiral anomaly}

In the Fujikawa analysis of the chiral anomaly in $D=4$ we encountered the quantity
\beq
\label{inden}
A(x) = \sum_n \phi_n^\dagger(x) \gbar \phi_n(x)
\eeq
where $\phi_n(x)$ are the eigenfunctions of $ i \Dslash$. Up to omission of the factor
of $\alpha(x)$, this is the trace of the matrix denoted by $C_{mn}$ in sec 1.3.

Now, if $\phi_n$ is an eigenfunction of $i \Dslash$ with non-zero eigenvalue $\lambda_n$, then
since $\gbar$ and $i \Dslash$ anti-commute, it follows that $\gbar \phi_n$ is an eigenfunction
with eigenvalue $- \lambda_n$. Thus, since eigenfunctions with different eigenvalues are orthogonal,
the integral of $A(x)$ only receives contributions from the eigenfunctions with zero eigenvalues,
\beq
\label{indenb}
\int d^4 x A(x) = \int d^4 x  \sum_i  (\phi_0^i (x))^\dagger \gbar (\phi_0^i(x))
\eeq
where $i$ labels the zero-modes. Now we can choose the zero modes to be eigenfunctions of
$\gbar$. Let $n_+$ be the number of zero-modes with eigenvalue $+1$ and $n_-$
the number of zero-modes with eigenvalue $-1$. Then since the non-zero modes cancel, the
right hand side of \calle{indenb} is formally $n_+ - n_-$, giving
\beq
\label{indenc}
\int d^4x A(x) = n_+ - n_-
\eeq

The problem with this argument is that it is purely formal, we are canceling off an infinite number
of zero terms. To make careful sense of this we need to regularize so that the sum converges, and
then take the limit as the regulator goes to infinity. This allows us to make the above argument precise,
but we also saw above that introducing the regulator allows us to compute $A(x)$ in terms of the
background gauge fields and yields a finite result when the regulator goes to infinity. Carrying this
out exactly as before we conclude that
\beq
\label{indend}
n_+ - n_- = {1 \over 32 \pi^2} \int d^4 x \epsilon^{\alpha \beta \mu \nu} F_{\alpha \beta}
F_{\mu \nu}
\eeq
Equation \calle{indend} is precisely the Atiyah-Singer index theorem for the Dirac operator coupled to
gauge fields in four dimensions.  Note that the quantity under the integral in \calle{indend}
is one half of the chiral anomaly \calle{zerobb}.  

Redoing this calculation on a $2n$-dimensional manifold
${\cal M}$ equipped with a spin connection $\omega$ with
curvature $R$ and for fermions in the representation ${\bf r}$ of
a $SU(N)$
gauge connection, $A$, with curvature $F$ gives the general
Atiyah-Singer index theorem:
\beq
\label{zn}
n_+ - n_- = \int_{\cal M} \left[
\hat A(R) ch(F) \right]_{2n} 
\eeq
where the $2n$ subscript indicates that we keep only the $2n$-form
part of the expression. Here the combination of Pontragin classes that contributes to the index, $\hat A(R)$, is the A-roof genus defined previously in \calle{zm}, and $ch(F)$ is the Chern character
given by
\beq
\label{zznn}
ch(F) = \Tr_{\br} e^{iF/ 2 \pi} = dim \br + c_1(F) + \cdots .
\eeq
The index density $\hat A(R) ch(F)$  is in general equal to one half of the chiral anomaly in $2n$ dimensions as
can be seen by keeping careful track of the factors of two in the Fujikawa analysis of the
chiral anomaly. As we will see in the following lecture, the index density in $2n$ dimensions
also provides the starting point for a derivation of the gauge anomaly in $2n-2$ dimensions.

\subsection{Exercises for Lecture 3}
\begin{itemize}
\item Consider an $SO(k)$ bundle $N$ with Pontragin classes $p_k(N)$.  In many physics
applications one also encounters the associated complex spin bundle $S(N)$ (for example, in
a gravitational theory fermion fields take values in $S(N)$). Compute the Chern classes
$c_2$ and $c_4$ for $S(N)$ in terms of the Pontragin classes $p_1(N)$ and $p_2(N)$. 
\item  Work out the gauge anomaly in $d=2,4$ by starting from the chiral anomaly in $d=4,6$
and applying the descent formalism. 
\end{itemize}

\section{Lecture 4: Anomaly inflow}
In the previous lecture we found that the descent procedure provides a candidate
for the gauge anomaly in $2n$ dimensions starting from the $2n+2$-form $\Tr F^{n+1}$
in $2n+2$ dimensions. From the results of the first lecture we also know that this 
$2n+2$-form is proportional to the chiral anomaly in $2n+2$ dimensions.

In this lecture we will explore a physical model which explains the
connection between gauge anomalies in $2n$ dimensions
and chiral anomalies in $2n+2$ dimensions
in more physical terms \cite{ch} and gives us the precise numerical factors
which relate the two anomalies. 
We will discover that there can be interactions in non-anomalous
theories in $2n+2$ dimensions which are anomalous in the presence
of $2n$-dimensional topological defects. The anomaly of the bulk 
interactions is
localized on the defect and expressed in terms of the chiral anomaly
in $2n+2$ dimensions via the descent procedure which encodes the
topology of the defect. The anomaly from these bulk interactions
is cancelled by an equal but opposite anomaly arising from fermion
zero modes localized on the defect.

The cancellation of anomalies between bulk terms and local terms
coming from zero modes on defects has many applications in string
theory and M theory, some of which we will discuss. There are also
interesting applications in condensed matter physics \cite{stone,kopnin} and lattice gauge theory \cite{kaplan}.

\subsection{Axion electrodynamics}

We will start by considering Dirac fermions in $D=4$ coupled vectorially
to a $U(1)$ gauge connection $A$ (which we will refer to as electromagnetism) and a complex scalar field $\Phi = \Phi_1
+ i \Phi_2$. We will later generalize to include coupling to gravity
through the spin connection $\omega$ and also to higher dimensions.
Since we are dealing with Dirac fermions the theory can be regulated
while maintaining gauge invariance (e.g. by Pauli-Villars regularization) and so
there are no gauge anomalies in this theory.

The Lagrangian is
\beq
\label{aaa}
{\cal L} = - {1 \over 4 e^2} F_{\mu \nu}F^{\mu \nu} +
\bar \psi i \Dslash \psi - \bar \psi (\Phi_1 + i \gbar \Phi_2) \psi
+ \partial_\mu \Phi^* \partial^\mu \Phi - V(|\Phi|^2)  
\eeq
We assume that the potential $V(|\Phi|^2)$ is chosen so that
$|\Phi|=v$ at the minimum for some non-zero $v$.

This theory has a global $U(1)$ Peccei-Quinn symmetry under which
the fields transform as
\beqn
\label{aab}
\Phi & \rightarrow & e^{i \alpha} \Phi,  \\
                  \psi & \rightarrow & e^{-i \alpha \gbar/2} \psi . 
\eeqn
Classically the chiral current $j_\mu^A$ associated to this symmetry
is conserved, but as we saw in the first lecture it is anomalous at
the quantum level with
\beq
\label{aac}
\partial_\mu j^\mu_A =
 { 1 \over 8 \pi^2} F^{\mu \nu} \tilde F_{\mu \nu} .
 \eeq

At the classical level there is a Nambu-Goldstone boson due to spontaneous
breaking of the $U(1)$ Peccei-Quinn symmetry. We can identify this field
by writing fluctuations of $\Phi$ about the vacuum as
$\Phi = v e^{ia}$ with $a$ the axion field. The axion couples to the
fermion fields through the interaction $\bar \psi e^{i a \gbar} \psi$.
This coupling can be removed by a chiral redefinition of the fermion fields,
but because of the anomaly  \calle{aac}  it reappears as a coupling to
$F \tilde F$. We can thus reduce the theory to a low-energy effective
theory of the photon and axion (axion QED) which will take the form
\beq
\label{aad}
S^{\rm axion-QED} = \int d^4 x \left( -{1 \over 4 e^2} F_{\mu \nu}
F^{\mu \nu} + {a \over 16 \pi^2} F_{\mu \nu} \tilde F^{\mu \nu}
+ {v^2 \over 2} \partial_\mu a \partial^\mu a - V(a) \right) 
\eeq

In the theory at hand the potential for the axion vanishes, $V(a) \equiv 0$.
In ``real world'' variants of this theory we would include the strong
interactions. It is then thought that QCD effects generate a potential
for the axion $V(a) \sim \Lambda_{QCD}^4 (1 - \cos{a})$.  For our purposes
we will ignore this complication. 

Because of the axion coupling to $ F \tilde F$, there are new sources
of the electromagnetic current when the axion field varies in spacetime.
Varying the action with respect to $A_\mu$ yields the equation of motion
\beq
\label{aae}
\partial^\mu F_{\mu \nu} = j_\nu,
\eeq
with
\beq
\label{aaf}
j_\nu = {e^2 \over 16 \pi^2} \partial^\mu
(a \tilde F_{\nu \mu} ). 
\eeq
Written in non-covariant form $j_\mu = (\rho, \vec j)$ we have
\beqn
\label{aag}
 \rho  & \sim  & a \vec \nabla \cdot \vec B +
\vec B \cdot \vec \nabla a , \\
\vec j & \sim  & \dot a \vec B + \vec \nabla a \times \vec E . 
\eeqn
The first term in $\rho$ is responsible for the Witten effect \cite{WittenEY}---monopoles
in the presence of an axion field or non-zero $\theta$ angle carry electric
charge---while the last term in $\vec j$ gives a Hall-like contribution
to the current which is perpendicular to the applied electric field.

Anomaly inflow arises by considering these contributions to the current
in the presence of topological defects. For $V(a)=0$ the vacuum manifold
has non-trivial $\pi_1$: $\pi_1({\cal M}_{\mbox{vac}}) = \IZ$ so the theory
has axion strings. For an axion string of charge $1$ the scalar field
$\Phi$ has the form
\beq
\label{aagh}
\Phi = f(\rho) e^{i \theta}
\eeq
where we are using polar coordinates with $z$ along the
string so that $\theta$ is the azimuthal angle and $\rho$ the radial
distance from the string. The function $f(\rho)$ approaches zero as
$\rho \rightarrow 0$ and $v$ as $\rho \rightarrow \infty$. The precise
form of $f(\rho)$ will not be needed in what follows.

Now consider the axion contributions to the electromagnetic current
in the presence of an axion string. If we apply an electric field
along the string, $\vec E = E \hat z$, then we find a current
$\vec j \sim -E \hat \rho/\rho$. The current is directed radially
inward and has a non-zero divergence on the string.  Clearly
electric charge can be conserved only if the axion string is capable
of carrying electric charge.

To understand how this can come about we should go back to the
Lagrangian describing the interaction of fermions with the axion
string configuration,
\beq
\label{aah}
{\cal L}_\psi = \bar \psi i \Dslash \psi + \bar \psi (
\Phi_1 + i \gbar \Phi_2 ) \psi .
\eeq
We can show that the string can carry charge by exhibiting normalizable zero modes of the
fermion equations of motion. 
To search for zero modes of the Dirac equation in the background \calle{aagh} we first
write $\gbar = \gamma^{\rm int} \gamma^{\rm ext}$ with $\gamma^{\rm int} = - \gamma^0 \gamma^1$
and $\gamma^{\rm ext} = -i \gamma^2 \gamma^3$. We also decompose the coordinates
into $x_{\rm int}^a = (x^0,x^1)$ and $x_{\rm ext} = (x^2,x^3)$.  Writing $\psi$ in terms
of eigenfunctions $\psi_\pm$ of $\gbar$ and looking for solutions independent
of $\phi$,  the Dirac equation becomes
\beqn
\label{aakl}
 i \gamma^a \partial_a \psi_- + i \gamma^2 ( \cos \theta + i \gamma^{\rm ext} \sin
\theta) \partial_{\rho} \psi_- & = &  f(\rho) e^{i \theta} \psi_+, \\
  i \gamma^a \partial_a \psi_+ + i \gamma^2 ( \cos \theta + i \gamma^{\rm ext} \sin
\theta) \partial_{\rho} \psi_+ & =  & f(\rho) e^{- i \theta} \psi_-,   
\eeqn
with solution
\beq
\label{aalm}
\psi_- = \eta(x_{\rm int}) \exp \left[ - \int_0^\rho f(\sigma) d \sigma \right] , \qquad
\psi_+ = -i \gamma^2 \psi_- , 
\eeq
with $i \gamma^a \partial_a \eta = 0$ and $\gamma^{\rm int} \eta = - \eta$.

Thus we see that the zero modes on the string are chiral and since they  couple to the
pullback of the spacetime gauge field, a $1+1$ dimensional observer
will conclude that the electromagnetic current on the string
is anomalous. Since the outside observer also sees an apparent violation
of current conservation we suspect that these two facts are related
and that in fact charge is conserved with an inflow of charge
from the outside of the correct magnitude to account for the anomaly
seen from the $1+1$ dimensional point of view.

It is actually more straightforward to demonstrate that the bulk
plus string action is gauge invariant than to show that the current
is properly conserved, although of course the two facts are related.
A proper understanding of current conversation requires a careful
study of the difference between the consistent and covariant forms
of the anomaly and may be found in \cite{nacu}.

\subsection{Anomaly inflow for axion strings}

We now want to show explicitly that the bulk and world-sheet contributions
to the anomaly cancel so that the overall theory is gauge invariant
in the presence of an axion string. That is,
we want to show that
\beq
\label{aai}
\delta_{\rm gauge} \left( S^{\rm axion-QED}_{\rm bulk} + S_{\rm string}
\right) = 0. 
\eeq
where $S_{\rm string}$ is the effective action for the axion string zero-modes.

To do this in a way which makes contact with our discussion in the previous
lecture, we first rewrite the relevant couplings in terms of characteristic classes.
The chiral anomaly in $D=4$ is given by
\beq
\label{aaj}
2 \omega_4  \equiv {1 \over 4 \pi^2}  F \wedge F ,
\eeq
where
\beq
\label{aaab}
\omega_4 = ch(F)|_4
\eeq
is the Dirac index density in four dimensions. Clearly $\omega_4$ is a characteristic class,
so we can apply the descent procedure.  An explicit calculation gives
\beqn
\label{aazs}
\omega_3^{(0)} & = & {1 \over 8 \pi^2} A \wedge F \\
\omega_2^{(1)} & = & {1 \over 8 \pi^2} \Lambda F 
\eeqn
where $\Lambda$ is the parameter of the gauge transformation, and we have used
the same notation for the descent procedure as in sec. 3.5. Now let us see how the
descent procedure appears in the analysis of the gauge variation of the bulk coupling of
the axion to gauge fields.

The bulk coupling of the axion to the gauge field given in \calle{aad} can be
written in terms of differential forms as 
\beq
\label{aakz}
\int_{M^4} {a \over 2} ~ 2 \omega_4  = \int_{M^4} a \omega_4.
\eeq
where the factor of $a/2$ arises because we need to do a chiral transformation
with parameter $\alpha= a/2$ in order to remove the coupling of the axion to the
fermion fields. We have also denoted the spacetime manifold by $M^4$.

Now in the presence of an axion string $a$ is not single valued
since it changes by $2 \pi$ in going around the string. Therefore \calle{aakz} doesn't
really make sense in the presence of an axion string. However,
$da$ \emph{is} single valued so we will integrate by parts and write the coupling \calle{aakz}
as
\beq
\label{aaoo}
 - \int_{M^4} da \wedge \omega_3^{(0)} .
\eeq
If we now vary \calle{aaoo} with respect to a gauge transformation we
have
\beq
\label{aap}
\delta \left[- \int_{M^4} da \wedge \omega_3^{(0)} \right]  = -  \int_{M^4} da \wedge
\delta \omega_3^{(0)}  = - \int_{M^4} da  \wedge d \omega_2^{(1)} 
= \int_{M^4} d^2 a  \wedge \omega_2^{(1)} 
\eeq

Now naively one might think that $d^2$ acting on any smooth function is zero and that
as a result \calle{aap} vanishes. However, $a$ is not a smooth function. It has winding number
one around the origin, just like the polar angle, and hence is not well defined at the origin. A more precise way
of saying this which is relevant to finishing the calculation of \calle{aap} is to consider
the integral of $da$ over any circle $S^1$ which encloses the axion string. Also, let $D^2$
be a disc whose boundary is this $S^1$. Then since $a$ has winding number one  and
using Stoke's theorem we have
\beq
\label{aal}
\int_{S^1} da = 2 \pi = \int_{D^2} d^2 a.
\eeq
Since this is true for any $S^1$ enclosing the origin (that is the axion string, at least
in the limit of an infinitely thin string), we must have
\beq
\label{aam}
d^2 a = 2 \pi \delta_2(\Sigma^2 \hookrightarrow M^4), 
\eeq
where $\delta_2( \Sigma^2 \hookrightarrow M^4)$ is a $2$-form delta
function with integral one over the directions transverse to the
string world-sheet $\Sigma^2$. In rectangular coordinates we would
have
\beq
\label{aan}
\delta_2(\Sigma^2 \hookrightarrow M^4) = \delta(x)\delta(y)
dx \wedge dy,
\eeq
and
\beq
\label{aann}
\int_{D^2} \delta_2 = 1.
\eeq
The reader might be wary of these manipulations of singular functions, and indeed
we will see later that this treatment of the string source is too naive for some purposes,
but for now this representation will suffice.

We can now finish the calculation of \calle{aap} using \calle{aam} to find
\beq
\label{aazxy}
\delta \left[- \int_{M^4} da \wedge \omega_3^{0)} \right]  = \int_{M^4} 2 \pi \delta_2(\Sigma^2 \hookrightarrow M^4) \wedge \omega_2^{(1)} = 2 \pi \int_{\Sigma^2} \omega_2^{(1)}.
\eeq
Thus the total action will be gauge invariant if the anomaly due
to the string zero modes is given by
\beq
\label{aaq}
- 2 \pi \int_{\Sigma^2} \omega_2^{(1) } .
\eeq
That is, if it is given by the descent procedure described in the previous lecture, starting with
the Dirac index density in four dimensions, and with a factor
of $2 \pi$ multiplying the final result. Since we know the theory must be consistent overall,
we could view this as a \emph{derivation}  that the two-dimensional gauge anomaly is given
by $2 \pi$ times the descent of the two higher-dimensional Dirac index density.

We can clearly generalize this construction to a theory with general non-Abelian anomalies
and to theories in higher dimensions. Thus, consider the Lagrangian \calle{aaa} in $2n+2$ dimensions
with  electromagnetism replaced by an arbitrary gauge group, and with the fermions in a representation
${\bf r}$ of the gauge group.  As before, we assume the
global Peccei-Quinn symmetry is spontaneously broken, and isolate the axion field
$a(x)$. In generalizing this construction to higher dimensions and particularly to superstring theory,
it is useful to introduce  the dual of the derivative of the axion field, $H_{2n+1} = {}^* d a$. The theory has axion ``strings,'' except that since we are now in $2n+2$ dimensions, these are actually $2n-1$-branes.
They couple to $B_{2n}$ with $H_{2n+1} = d B_{2n}$ such that 
\beq
\label{aax}
d {}^* H_{2n+1} = 2 \pi \delta_2 (\Sigma^{2n} \hookrightarrow M^{2n+2}).
\eeq

This generalization has non-chiral fermions in $2n+2$ dimensions, but one can verify
that on the axion $(2n-1)$-brane there are chiral fermions in the same representation ${\bf r}$
of the gauge group as the bulk fermions. These fermions have an anomaly which  must be canceled
by inflow from the bulk.

The bulk inflow is provided by a coupling which arises as before, by using the chiral anomaly
of the bulk theory to integrate out the bulk fermions which leaves one with the bulk coupling
\beq
\label{aay}
-  \int_{M^{2n+2}} {}^* H_{2n+1} \wedge \left( ch(F) \right)^{(0)}_{2n+1} .
 \eeq
Here we have introduced a generalization of our previous notation where the superscript
$(0)$ means the form one obtains by applying the first step of the descent procedure to
the quantity in brackets (in this case the $2n+2$-form part of $ch(F)$). 
Working out the gauge variation of this term in the presence of the $2n-1$ brane as before
shows that the inflow precisely cancels the zero mode anomaly provided that the zero mode
anomaly is given by the descent procedure just discussed.

\subsection{Gravitational anomaly cancellation}

One interesting generalization of the previous results arises when we
couple the theory to gravity. In doing this we will encounter
subtleties which presage some of the problems which arise in the
theory of fivebranes in string theory and M theory. This analysis also provides a
nice application of some of the formalism we developed in lecture 3 involving the
manipulation of characteristic classes.

The gravitational contribution to the Dirac index density in $D=4$ is  the $4$-form
part of $\hat A (R)$ which from \calle{zm} is $- p_1(R)/24$.  There will thus be a coupling
to the axion of the form \calle{aakz} but with 
\beq
\label{aa}
\omega_4 = {1 \over 8 \pi^2} F \wedge F - {p_1(R) \over 24}.
\eeq

To figure out the implications of this new term for anomaly cancellation, we first need to ask
about the symmetries of the axion string configuration.  Of the $SO(3,1)$ local Lorentz symmetry,
only $SO(1,1) \times SO(2)$ leaves the axion string configuration invariant. Thus we should check
that these symmetries are not anomalous. Geometrically, $SO(1,1)$ transformations act
on the tangent bundle to the string world-sheet, $T \Sigma^2$, while the $SO(2)$ transformations
act as gauge transformations on $N$, the normal bundle to the string world-sheet. Corresponding
to this decomposition, we can decompose the tangent bundle restricted to the string
world-sheet as
\beq
\label{aau}
 TM|_{\Sigma^2} = T \Sigma^2 \oplus N.
\eeq
We can then use \calle{zjjj} to deduce that $p_1(TM) = p_1(T\Sigma^2) + p_1(N)$. And, following
through the rest of the calculation for anomaly inflow for the axion string, we see that
there is a gravitational inflow contribution to the anomaly given by the descent of
\beq
\label{aaw}
I^{\rm inflow} = -  \left( p_1(T \Sigma^2) + p_1(N) \right) /24 .
\eeq

Now let us compare this to the anomaly of the chiral zero modes, again focusing only
on the gravitational contribution. The normal bundle $N$ has an $SO(2)$ structure group.
The curvature can be represented by a $2 \times 2$ antisymmetric matrix. Let the skew eigenvalues
be $\pm x$.  Thus $p_1(N) = x^2$. However, the fermion zero modes transform in the spinor
representation of $SO(2)$, $S(N)$, so the curvature in this complex representation has eigenvalues
$\pm x/2$. The total gravitational anomaly of the fermion zero modes is thus given by descent of
\beq
\label{aaxx}
 - {1 \over 2} ch S(N) \hat A (T \Sigma^2) |_4.
 \eeq
From the discussion above we have
\beq
\label{aayy}
ch S(N) = e^{x/2} + e^{-x/2} = 2 + x^2/4 + \cdots = 2 + {p_1(N) \over 4} + \cdots 
\eeq
which gives for the total anomaly
\beq
\label{aaz}
I^{\rm zero mode} =  {p_1(T \Sigma^2) \over 24} - {p_1(N) \over 8} .
\eeq
Adding \calle{aaw} and \calle{aaz} gives
\beq
\label{aazz}
I^{\rm total}= I^{\rm inflow} + I^{\rm zero mode} = - {p_1(N) \over 6}. 
\eeq

So, while the tangent bundle anomaly cancels, it appears that the normal bundle anomaly
does not!  The resolution of this puzzle was first pointed out in \cite{wittfive} in an analysis of
fivebrane anomalies in IIA string theory. One important ingredient is to note that since $N$ is an even-dimensional
bundle, the top Pontrajin class can be factorized in terms of the Euler class,
$p_1(N) = e^2(N)$ with $e(N)$ the Euler class. We can thus write the uncanceled anomaly
as $- e^2(N)/6$.  

A second key ingredient to canceling the normal bundle anomaly is to realize that
the definition of $\delta_2(\Sigma^2 \hookrightarrow M^4)$ requires modification in a theory
that includes gravity. The modifications are somewhat complicated and will be  discussed in detail
in the following lecture. For now, the only fact we will need is that the modifications ensure
that the connection $A$ on the normal bundle is such that
\beq
\label{aaxyz}
d H_1|_{\Sigma^2} = \delta_2|_{\Sigma^2} = e(F) ,
\eeq
where $e(F)$ is the representative of the Euler class for the connection $A$ with curvature $F$
which is induced from the spin-connection in spacetime.

As a result of these two facts, we see that there is a local counterterm we can add to the
effective action on the axion string which will cancel the normal bundle anomaly. Namely,
we add the term
\beq
\label{abxy}
\int_{\Sigma^2} {1 \over 6}  H_1|_{\Sigma^2} \wedge (e(F))_1^{(0)}.
\eeq
Computing the variation of this term under gauge transformations of the normal bundle
and using \calle{aaxyz} we see that its variation precisely cancels the anomaly \calle{aazz}.

\subsection{Anomalous couplings on D-branes}

Another nice example of the inflow mechanism occurs in the study of Chern-Simons or anomalous
couplings on D-branes \cite{ghm,cy,mm}.  Consider type
II string theory and define a formal sum of RR potentials as
\beqn
\label{threea}
C & = & C_1+C_3 + \cdots \qquad IIA \\
                                       C & = & C_0+C_2 + \cdots \qquad IIB 
 \eeqn
In the absence of D-branes, the corresponding field strength is $H=dC$. 
On a D$p$-brane with worldvolume $B^{p+1}$ there is a coupling of $C$ to
anomalous or Chern-Simons terms given by
\beq
\label{threeb}
\int_{B^{p+1}} C \wedge ch(F) {\sqrt{\hat A(TB^{p+1})} \over \sqrt{\hat A(N B^{p+1})}}
\eeq
This coupling was deduced using anomaly inflow arguments \cite{ghm,cy,mm} and
played an important role in suggesting the importance of K-theory in the classification
of D-brane charges \cite{mm,Wittenk}. 

We will discuss part of this coupling by considering the special case of
two  D5-branes in IIB string theory which intersect along a 1-brane,
$D5_1 \cap D5_2 = I1$.    For example we can
take the $D5_1$ worldvolume $\Sigma_1^6$ to lie along the $0,1,2,3,4,5$ directions, and the $D5_2$ worldvolume $\Sigma_2^6$ to
lie along the $0,1,6,7,8,9$ directions and then the $I1$ brane at the intersection has its worldvolume
$\Sigma^2$ along $0,1$.
We will also focus only on the terms involving the gauge field,
leaving the generalization to the gravitational couplings as an exercise. 
Assume that $D5_{1,2}$ has Chan-Paton labels $N_{1,2}$
so that there is a $U(N_{1,2})$ gauge group on $D5_{1,2}$.
The zero modes that are localized on the intersection come from open string which start
on one $D5$ and end on the other. They thus transform in the $(\bar N_1,N_2) +
(N_1,\bar N_2)$ representation of the $U(N_1) \times U(N_2)$ gauge group.  By working through the standard quantization of open strings, or by thinking about the zero modes resulting from the
supersymmetries broken by the intersection, one sees that these zero modes are chiral on the
$I1$ worldvolume. They thus have a gauge anomaly which is determined by descent from
the $4$-form
\beqn
\label{threec}
I^{\rm zero modes} & = & {1 \over 2}  \left( ch_{(\bar N_1,N_2)}(F_1)+ch_{(N_1,\bar N_2)}(F_2) \right) \bigm |_4 \\
& = &  ch_{N_1}(F_1) ch_{N_2}(F_2) \bigm |_4  \\
& =  &  N_1 c_2(F_2) +  N_2 c_2(F_1) +  c_1(F_1) c_1(F_2). 
 \eeqn
The factor of $1/2$ in \calle{threec} accounts for the reality of the fermion representation.

Since there are no gauge fields in the ten-dimensional bulk, the anomaly \calle{threec} must be cancelled
by inflow from the $D5$-branes onto the $I1$-brane at the intersection. 
To see how this happens we study the possible anomalous couplings of the RR
potentials to gauge fields.  We assume that the  D5-branes have couplings of the
form
\beq
\label{threeda}
S_{\rm anom} =  \sum_i  \left[ - {1 \over 2} \int_{\Sigma_i^6} \left( N_i C  -  H \wedge Y_{i}^{(0)}(F_i) \right)
\right]
\eeq
where $i=1,2$ labels the two $D5$-branes. 

Some words of explanation are in order regarding this ansatz. The first term in \calle{threeda}
simply expresses the fact that the $D5$-brane acts as a source for $C$ ($C_6$ to be precise)
with strength $N_i$. The second term involves  a characteristic form $Y(F) = d Y(F)^{(0)}$ and
an integration by parts as compared to this
term written in terms of $C$ for the same reason as in \calle{aaoo}: $C$ is not single-valued in
the presence of D-branes. Finally, the factor of $1/2$
accounts for the fact that this action is written in terms of both ``electric'' and ``magnetic"
potentials. See \cite{cy} for details. 

The action \calle{threeda} leads to the equations of motion/Bianchi identities
\beq
\label{threedb}
d H = - \delta_4(\Sigma_1^6 \hookrightarrow M^{10}) \wedge Y_1^{(0)}(F_1) - \delta_4(\Sigma_2^6
\hookrightarrow M^{10}) \wedge Y_2^{(0)}(F_2)
\eeq
Gauge invariance of $H$ then requires that $C$ vary under gauge transformations as
\beq
\label{threedc}
\delta C = \delta_4(\Sigma_1^6  \hookrightarrow M^{10} ) \wedge Y_1^{(1)} + \delta_4(\Sigma_2^6
\hookrightarrow M^{10} ) \wedge Y_2^{(1)}.
\eeq

Using this, we can compute the gauge variation of \calle{threeda}:
\beqn
\label{threedd}
\delta S_{\rm anom} & = & -  \int_{M^{10}} \delta_4 (\Sigma_1^6 \hookrightarrow M^{10})
\wedge \delta_4(\Sigma_2^6 \hookrightarrow M^{10}) \wedge (Y_1 \wedge Y_2 )^{(1)} \\
& = & - \int_{\Sigma^2} ( Y_1 \wedge Y_2)^{(1)} 
\eeqn
We thus see that the anomaly will cancel between \calle{threedd} and \calle{threec}
provided that $Y(F) = ch (F)$.

By doing such an analysis systematically and including tangent and normal bundle
gravitational anomalies one can derive the full set of couplings given in \calle{threeb}.
For details see \cite{ghm,cy,mm}. The dynamics of this system and the role played by anomaly
inflow have recently been analyzed in great detail in \cite{iks}.

\subsection{Exercises for Lecture 4.}

\begin{itemize}
\item  Verify that \calle{aalm} is a normalizable solution to \calle{aakl}.
\item  Compute the gravitational contribution to the anomaly of the zero modes on the
intersection of two $D5$-branes, and verify that the anomaly is cancelled by inflow
from the interaction \calle{threeb}.
\end{itemize}

\section{Lecture 5: M5-brane anomalies}

Since M theory is thought to be the mother theory from which all
string theories arise, it is natural to study the cancellation of
anomalies for the extended objects of M theory. M
theory is known to have two types of BPS extended objects,
membranes and fivebranes which we will denote as M2
and M5. Since the M2-brane has an odd-dimensional
world-volume it does not have anomalies in continuous symmetries.
There is a parity anomaly which is connected  to the quantization condition of the $4$-form
field strength of M theory \cite{wittflux}.

The M5 brane is a more interesting and subtle object. For charge
$Q_5=1$ it has zero modes which comprise a tensor multiplet of
six-dimensional $(2,0)$ supersymmetry \cite{chs,km}. The free field theory
of these zero modes has a $Spin(5)_R$ symmetry and
the tensor multiplet
contains a $Spin(5)_R$ singlet, 
a $2$-form $B_2^+$ with self-dual field strength,
$H_3 = d B_2^+ = {}^* H_3$ , chiral fermions $\psi$ transforming in the
spinor representation of $Spin(5)_R$,
and five scalar fields transforming as the vector of $Spin(5)_R$.

For fivebrane charge $Q_5>1$ the M5-brane is not well understood.
A first principles definition of the theory does not exist, although
there are partial results on the primary conformal fields and correlations
functions which follow from a Matrix theory formulation of the theory
and from application of the AdS/CFT correspondence  to the $AdS_7 \times S^4$
near horizon geometry of the M5-brane. We will see that anomaly inflow
yields some additional information about this theory.

So far we have only discussed anomalies for chiral spin $1/2$ fermions.
It is clear however that there are potential anomalies for other
chiral fields. For example, in $1+1$ dimensions we can have chiral
scalars $\phi$ obeying $d \phi = \pm {}^* d \phi$ which are equivalent
upon fermionization to chiral fermions and therefore must contribute
to the gravitational anomaly. Similarly, in $D=6,10$ one can have
bosonic $2$-form and $4$-form  potentials with self-dual or anti-self-dual field strengths. The duality constraint means that the fields
transform chirally under the Lorentz group. There is no way to regulate
a theory with such fields without violating Lorentz invariance, so
there is a potential gravitational anomaly. These gravitational anomalies
were analyzed in \cite{agw} and we will simply apply their results here
without further discussion.

\subsection{Tangent bundle anomalies and bulk couplings}

For a charge $Q_5=1$ M5-brane the gravitational anomaly on the M5
has contributions from the fermions and the self-dual $2$-form. Since the world-volume $W^6$ is
six-dimensional, the descent formalism implies that both
contributions are summarized as in our previous discussions by an
$8$-form characteristic class. In this section we will
focus only on the tangent bundle anomaly. The normal bundle anomaly
will require a considerably more complicated treatment.
The fermion contribution to the tangent bundle anomaly is given
by
\beq
\label{baa}
I_8^{\rm ferm} = 2 \hat A(TW^6)|_{\rm 8-form} =
{1 \over 5760} \left( 14 p_1^2(TW^6) - 8 p_2(TW^6) \right) 
\eeq
The factor of two in front of $\hat A$ arises because there are four fermions (transforming
as a {\bf 4} of $Spin(5)_R$), but they obey a Majorana constraint reflecting their origin
as Majorana spinors in eleven dimensions which reduces the anomaly by a factor of $1/2$.

It follows from the results of \cite{agw} that the $2$-form contribution
is given by
\beq
\label{bab}
I_8^{B^+} = {1 \over 5760} \left( 16 p_1^2(TW^6) - 112 p_2(TW^6).
\right) 
\eeq

The total tangent bundle anomaly is thus
\beq
\label{bac}
I_8^{\rm total} = I_8^{\rm ferm} + I_8^{B^+} =
{1 \over 192} \left( p_1^2(TW^6) - 4 p_2(TW^6). \right) 
\eeq
This anomaly must be cancelled by inflow from the bulk in a way
which is quite analogous to what happens for the axion string. The
M5-brane acts as a magnetic source of $C_3$ via
\beq
\label{bad}
d G_4 = 2 \pi \delta_5 (W^6 \hookrightarrow M^{11}) 
\eeq
with $G_4$ the field strength for $C_3$.  We can therefore cancel
the anomaly via inflow \cite{Duff} if there is a bulk coupling given by 
\beq
\label{bae}
\int_{M^{11}} C_3 \wedge X_8
\eeq
with
\beq
\label{baf}
X_8 = - {1 \over 192} \left( p_1^2(TM^{11}) - 4 p_2(TM^{11})
\right).
\eeq
The verification that this cancels the tangent bundle anomaly is completely analogous
to the demonstration for the axion string and is left as a small
exercise for the reader.

\subsection{The normal bundle anomaly}

Although the previous section shows how the tangent bundle anomaly
cancels between the world-brane and bulk contributions, as in our
discussion of the axion string, we still need to analyze possible
anomalies in diffeomorphisms which act as $SO(5)$ gauge transformations
on the normal bundle as well as mixed tangent bundle-normal
bundle anomalies.  The M5-brane background breaks the $D=11$ Lorentz
symmetry $Spin(10,1)$ to $Spin(5,1) \otimes Spin(5)$ and corresponding
to this we can decompose the restriction of the spacetime tangent bundle
to the fivebrane world-volume as
\beq
\label{bag}
T M^{11}|_{W^6} = T W^6 \oplus N 
\eeq
with $N$ the normal bundle. The fact that there are diffeomorphisms
acting as $SO(5)$ gauge transformations may be more familiar to some
readers in the context of the AdS/CFT correspondence \cite{magoo}. In the near horizon
limit the M5 geometry becomes $AdS_7 \times S^4$ which can be viewed
as an $S^4$ Kaluza-Klein compactification of $D=11$ supergravity. The
resulting supergravity on $AdS_7$  has an $SO(5)$ gauge group coming
from the isometry group of $S^4$.

There are two obvious contributions to the normal bundle anomaly coming
from the M5-brane zero modes and via inflow from the bulk term
\calle{bae} determined above by cancellation of the tangent bundle anomaly.

The antisymmetric tensor field does not contribute to the normal
bundle anomaly since it is a singlet under $SO(5)$. The fermion
fields transform as a $\bf 4$ under $Spin(5)$, that is they take
values in the rank four spin bundle $S(N)$. The total anomaly due
to the fermion fields is thus the $8$-form part of
\beq
\label{bah}
I_8^{\rm ferm} ={1 \over 2} ch S(N) \hat A(TW^6)|_{8} 
\eeq
It is useful to represent the Chern classes of $S(N)$ in terms of
Pontrajin classes. This can be done by writing the curvature of the normal
bundle as in \calle{zq} and noting that the eigenvalues of the curvature of
$S(N)$ will then be $\pm (x_1+x_2)/2$, $\pm (x_1-x_2)/2$. Using this and
the definition of $ch(F)$ from \calle{zznn} gives
\beqn
\label{bai}
ch S(N) & = & e^{(x_1+x_2)/2}+e^{(x_1-x_2)/2}+
           e^{(-x_1+x_2)/2}  + e^{(-x_1-x_2)/2}    \\
                         & = & 4 + (x_1^2+x_2^2)/2 + (x_1^4+x_2^4 +
                               6 x_1^2 x_2^2)/96 + \cdots   \\
                         & = & 4 + {p_1(N) \over 2} +
                         {p_1(N)^2 \over 96} + {p_2(N) \over 24} +
                         \cdots 
\eeqn

The contribution from the bulk term is easily computed using
\beq
\label{beej}
p_1(TM^{11}|_{W^6}) = p_1(TW^6)+p_1(N)
\eeq
and
\beq
\label{bbek}
p_2(TM^{11}|_{W^6}) = p_2(TW^6)
+ p_2(N) + p_1(TW^6) p_1(N)
\eeq
to give
\beq
\label{baj}
I_8^{\rm inflow} = -{1 \over 48} \left( {p_1(TW^6)^2 + p_1(N)^2
- 2 p_1(TW^6) p_1(N) \over 4} - p_2(TW^6) - p_2(N) \right)
\eeq

Adding \calle{baj} and \calle{bah} using \calle{bai} gives for the total anomaly
\beq
\label{bak}
 I_8^{\rm ferm} + I_8^{\rm inflow} = {p_2(N) \over 24} .
\eeq

So as in the axion string case, there is a normal bundle anomaly
which does not cancel between the zero modes and bulk terms. However
unlike the axion string case, here the solution is not so clear. Note
that if we were considering the IIA fivebrane in $D=10$ then the
normal bundle would be a $SO(4)$ bundle, $p_2(N)$ would be the square
of the Euler class of this bundle, and the anomaly could be cancelled
by adding a term to the fivebrane worldvolume in analogy to what we
did for the axion string as was first shown in \cite{wittfive}. The relationship
between the anomaly cancellation discussed below for the M5-brane and that
for the IIA five-brane can be found in \cite{Beckers}. 

This will not work for the M5-brane. The Euler class vanishes for
odd rank $SO(N)$ bundles and there is simply no way to factorize
the uncancelled anomaly. Thus something new is required. What this
something new is, is not a priori obvious. One might think that since
the extremal M5-brane is a smooth solution to $D=11$ supergravity,
we should try to study the diffeomorphism invariance directly in supergravity. This
would involve things like the study of the 
Rarita-Schwinger operator in the M5-brane background geometry. However we have seen that
the study of anomalies really requires a study of families of backgrounds,
and perturbations of the M5-brane geometry are generically singular. Thus studying the
Rarita-Schwinger  operator in a sufficiently general background is a daunting task.
To my knowledge no serious attempts have been made to study the problem
this way.  One might also hope to address the problem in Matrix Theory
since it claims to be a fundamental formulation of M theory \cite{bfss}. It is clear
though that this is not practical with current technology. We don't even
have a proof of $D=11$ Lorentz invariance in Matrix Theory so we are
ceratinly not in a position to be looking for quantum violations of
local Lorentz invariance.

There is a formalism which allows one to cancel the remaining normal
bundle anomaly which we will now discuss. It involves a more careful treatment of the M5-brane
source term and the structure of $D=11$ supergravity in the presence
of fivebranes. It probably should not be viewed as a final understanding
of the problem.  One would eventually hope for a microscopic formulation
of M theory which makes some of the  manipulations in the following section appear more
natural.

\subsection{Bump forms and Thom classes}

So far in our discussion of M5-brane anomalies, and in fact for anomalies on all branes,
we have treated the branes as singular sources.  Thus we have treated the
M5-brane as a singular magnetic source for the $3$-form field
with
\beq
\label{bal}
d G_4 = 2 \pi \delta_5(W^6 \hookrightarrow M^{11}) .
\eeq
In our discussion of normal bundle anomalies for the axion string
we already saw that this was not completely correct, or at least
not completely well specified without a more precise definition of 
$\delta_5$, because the source
term must depend in a non-trivial way on the connection on
the normal bundle.   In addition, we might expect that singular
sources might be problematic in situations where the equations of
motion are non-linear. In M theory this occurs even for the $3$-form
field alone due to the Chern-Simons like term in the action
\beq
\label{bam}
S_{CS} = - {2 \pi \over 6} \int_{M^{11}} {C_3 \over 2 \pi}
\wedge {G_4 \over 2 \pi} \wedge {G_4 \over 2 \pi} .
\eeq

In this section we will describe a method of smoothing out the
source term and including the dependence on the connection on the
normal bundle. This will involve a very lowbrow presentation of
mathematics which is covered more rigorously but still in an
accessible way in \cite{bt}. We will see that it also becomes necessary
to modify
the term \calle{bam} in the presence of an M5-brane, or it might be more precise
to say, give a careful definition of \calle{bam} in the presence of a M5-brane. In the
following section we show that this modification or definition cancels the
remaining anomaly \calle{bak}.

To begin with we  will make sure that bulk integrals in the presence of
an M5-brane are well defined. We do this by cutting out a region of
radius $\epsilon$ around the M5-brane.
At each point on $W^6$ we thus have a disk of radius $\epsilon$. We will denote
the total space of this disk bundle over $W^6$ by $D_\epsilon(W^6)$.
To define bulk integrals we cut out this ``tubular neighborhood'' of the 
fivebrane and then take the limit as $\epsilon \rightarrow 0$:
\beq
\label{ban}
\int_{M^{11}} {\cal L}_{bulk} \equiv \lim_{\epsilon \rightarrow 0}
\int_{M^{11} - D_\epsilon(W^6)} {\cal L}_{bulk} 
\eeq
Later on we will see that certain bulk terms are boundary terms and we
will then use the fact that the boundary of $D_\epsilon(W^6)$ is
$S_\epsilon(W^6)$, the sphere bundle of radius $\epsilon$ over $W^6$.

Next we smooth out the fivebrane source. The anomaly should be insensitive
to the precise profile of the fivebrane, so we introduce a function
of the radial direction away from the fivebrane, $\rho(r)$ with
$\rho \rightarrow -1$ as $r \rightarrow 0$ and with $\rho$ vanishing
for $r$ larger than some finite value. The derivative $d \rho$ is called
a bump form and integrates to $1$ in the radial direction, $\int d \rho=1$.

If we were considering a fivebrane in ``flat space'' with a vanishing
connection on the normal bundle then we could write the smoothed out
source term as
\beq
\label{bao}
\delta_5(W^6 \rightarrow M^{11}) \sim  d \rho  \epsilon_{
a_1 \cdots a_5} d \hat y^{a_1} \cdots d \hat y^{a_4} \hat y^{a_5} 
\eeq
where $\hat y^a = y^a/r$ are isotropic coordinates normal to the
M5-brane.  We choose the proportionality constant in \calle{bao} so that the r.h.s. has
integral one over the space transverse to the fivebrane. This can then be viewed as an
approximation to a delta function by choosing $\rho$ to be supported in an arbitrarily small
neighborhood of the origin.

To study normal bundle anomalies we need to vary the
normal bundle connection and we need a generalization of \calle{bao}
which is properly covariant under $SO(5)$ gauge transformations of
the normal bundle.
We will write the fivebrane source equation as
\beq
\label{bap}
d (G_4/2 \pi) = d \rho \wedge e_4/2
\eeq
and determine $e_4$ by demanding that it obey the following
conditions.
\begin{itemize}
\item Taking the exterior derivative of \calle{bap} implies that
$d e_4 = 0$.
\item  $e_4$ should be covariant under $SO(5)$ gauge transformations.
\item  The integral of $e_4/2$ over the fibres of $S_\epsilon(W^6)$
should equal one in order that we have the correctly normalized fivebrane
charge.
\item  The expression for $e_4/2$ should be proportional to
$\epsilon_{a_1 \cdots a_5} d \hat y^{a_1} \cdots d \hat y^{a_4}
 \hat y^{a_5}$ when the $SO(5)$ connection on $N$ is trivial.
\end{itemize}

To make $e_4$ covariant it is natural to replace ordinary derivatives
by covariant derivatives. Recall that for a principal fibre bundle
with connection $\Theta^{ab}$ we can split the tangent space at a point
$x$ into vertical and horizontal components $T_x = V_x \otimes H_x$
with basis $(\partial/\partial \hat y^a, D_\mu = \partial/\partial x^\mu
- \Theta_\mu^{ab} \hat y^a \partial/\partial \hat y^b ) $.

To construct
$e_4$ we want to generalize $d \hat y^a$ to the $1$-form $(D \hat y)^a$
in $T_x^*$ which is dual to $\partial/\partial \hat y^b$, that is which
satisfies
\beq
\label{baar}
 \langle (D \hat y)^a, D_\mu \rangle =0, \qquad
\langle (D \hat y)^a, \partial/\partial \hat y^b \rangle = \delta^{ab} .
\eeq
This gives $(D \hat y)^a = d \hat y^a - \Theta^{ab} \hat y^b$.

It thus seems natural to guess that
\beq
\label{bas}
e_4 \sim \epsilon_{a_1 \cdots a_5} (D \hat y)^{a_1} \cdots
(D \hat y)^{a_4} \hat y^{a_5} .
\eeq
However this is not correct since it is easy to see that \calle{bas} is not
closed using
\beq
\label{baat}
d (D \hat y)^a = \Theta^{ab} (D \hat y)^b - F^{ab} \hat y^b 
\eeq
where $F^{ab} = d \Theta^{ab}  - \Theta^{ac} \wedge \Theta^{cb}$ is
the field strength of $\Theta$. Nonetheless it is possible to add
additional terms to cancel these unwanted terms and to construct a form for
$e_4$ which is both covariant and closed. This leads to\footnote{This corrects \cite{Beckers}
the expression given in \cite{fhmm} by a factor of two.}
\beqn
\label{bau}
e_4 & =  &{1 \over 32 \pi^2} \epsilon_{a_1 \cdots a_5} [
 (D \hat y)^{a_1} \cdots (D \hat y)^{a_4} \hat y^{a_5}  \\
 & -  & 2 F^{a_1 a_2}
 (D \hat y)^{a_3} (D \hat y)^{a_4} \hat y^{a_5}  
  +  F^{a_1 a_2}
 F^{a_3 a_4} \hat y^{a_5} ]  
 \eeqn

This expression for $e_4$ satisfies all the conditions above so we take
as our smoothed out, covariant fivebrane source $d \rho \wedge e_4/2$.
The quantity $e_4/2$ is known in the mathematics literature as the global
angular form and the source term $d \rho \wedge e_4/2$ is a particular
representative of the Thom class of the normal bundle.

\subsection{Anomaly cancellation}

We have found that the properly defined fivebrane source has
metric dependence. It particular it will vary under diffeomorphisms
which act as $SO(5)$ gauge transformations on the normal bundle.
We can derive this variation by applying the descent procedure to
$e_4$ since it is a characteristic class. We thus have
\beqn
\label{bav}
e_4  & = & d e_3^{(0)} \\
                  \delta e_3^{(0)} & = &  d e_2^{(1)} 
\eeqn

We can then reexpress the uncancelled normal bundle anomaly by
using a result of Bott and Cattaneo \cite{bc}:
\beq
\label{baw}
\int_{S_\epsilon(W^6)} e_4 \wedge e_4 \wedge e_2^{(1)}
= 2 \int_{W^6} p_2^{(1)}(N)  .
\eeq

The left hand side of \calle{baw} gives a cubic form for the uncanceled anomaly which is 
reminiscent of the Chern-Simons coupling
of $D=11$ supergravity
\beq
\label{bax}
S_{CS} = - {2 \pi \over 6} \int_{M^{11}} (C_3/2 \pi) \wedge d (C_3/2 \pi)
\wedge d (C_3/2 \pi).
\eeq
and so suggests that we think carefully about the proper definition of this term in the
presence of M5-branes.

In the absence of M5-branes we have $G_4=dC_3$ but in the presence of
M5-branes we have found a smoothed out source equation
\beq
\label{baay}
d (G_4/2 \pi) = d \rho \wedge e_4/2
\eeq
which is not compatible with $G_4 = dC_3$. We can integrate \calle{baay} to
find
\beq
\label{baz}
(\Gslash_4/2 \pi) = d( \Cslash_3/2 \pi)  + A \rho e_4/2 - B d \rho e_3^{(0)}/2
\eeq
with $A+B=1$. Here $C_3$ describes flucutations about the M5-brane background.
Since we have smoothed out the source, physical quantities like $G_4$
should be smooth at the origin. This requires that $A=0$ since $e_4$
is singular at $r=0$ much like $d \theta$ is singular at the origin
in polar coordinates. So in the presence of a M5-brane the correct
relation between $G_4$ and $C_3$ is
\beq
\label{bazz}
(\Gslash_4/2 \pi) = d ( \Cslash_3/2 \pi) - d \rho \wedge e_3^{(0)}/2 .
\eeq

This is roughly analogous to the relation $H_3 = d B_2 - \omega_3$
which occurs in the low-energy theory of the heterotic string and
which plays an important role in the Green-Schwarz anomaly cancellation
mechanism \cite{greenschwarz}. In particular, for $G_4$ to be invariant under $SO(5)$ gauge
transformations,  $C_3$ must vary according to
\beq
\label{bayy}
\delta ( \Cslash_3/2 \pi)  = - d \rho \wedge e_2^{(1)}/2
\eeq
and this implies the descent relations
\beqn
\label{baxx}
(\Gslash_4/2 \pi) - \rho e_4/2 & =  & d (( \Cslash_3/2 \pi) - \sigma_3) \\
 \delta (( \Cslash_3/2 \pi) - \sigma_3 ) & = & - d(\rho e_2^{(1)}/2) 
 \eeqn
where we have defined $\sigma_3 \equiv \rho e_3^{(0)}/2$.

We now have to decide what the proper form of \calle{bax} should be in the
presence of a fivebrane. There is clearly some ambiguity since $dC_3$
and $G_4$ differ in the presence of a fivebrane. A natural choice which
also cancels the anomaly is to preserve the Chern-Simons form of \calle{bax}, that
is we try to generalize \calle{bax} in the presence of a fivebrane so that it
still has the form $\int x \wedge dx \wedge dx$ for some $3$-form $x$.
Using the descent relation \calle{baxx} suggests that the correct form for the
Chern-Simons coupling is
\beq
\label{baww}
S'_{CS}= \lim_{\epsilon \rightarrow 0}  - {2 \pi \over 6}
 \int_{M^{11}-D_\epsilon(W^6)} ((\Cslash_3/2 \pi) - \sigma_3) \wedge
d((\Cslash_3/2 \pi) - \sigma_3) \wedge d ((\Cslash_3/2 \pi) - \sigma_3) 
\eeq
where we have also recalled the definition of bulk integrals \calle{ban}.

We can now compute the variation of the modified Chern-Simons term under
diffeomorphsims acting as gauge transformations on the $SO(5)$ normal
bundle. Using \calle{baxx} we have
\beqn
\label{bavv}
\delta S'_{CS} & =  & \lim_{\epsilon \rightarrow 0}
\int_{M^{11}-D_\epsilon(W^6)} {2 \pi \over 6} d(\rho e_2^{(1)}/2) \wedge
((G_4/2 \pi) - \rho e_4/2)^2 \\
& =  & -  \lim_{\epsilon \rightarrow 0} \int_{S_\epsilon(W^6)} {2 \pi \over 6}
\rho {e_2^{(1)} \over 2} \wedge \rho {e_4 \over 2} \wedge \rho {e_4 \over
2} + O(\epsilon) \\
&=  &  2 \pi \int_{W^6} {p_2^{(1)}(N) \over 24} 
\eeqn
where we have integrated by parts, used the fact that $G_4$ is smooth near
the fivebrane, and applied the Bott-Catteneo formula \calle{baw}.

We thus  find that the normal bundle anomaly does indeed cancel, but
only after properly including the normal bundle dependence in the
fivebrane source and modifying the supergravity action in the presence
of a fivebrane.

Although this mechanism must be correct in some sense given the intricate
way that the different pieces fit together, it raises as many questions
as it answers. First, the description we have given is in the spirit
of a low-energy effective action. It would be nice to have a microscopic
derivation of the modified source terms and Chern-Simons couplings. For a treatment
of these questions in simpler models see \cite{hr,bhr}. These modifications
are of course not unique, and some other forms have been proposed which also
cancel the anomaly \cite{Boyarsky}. Some hints at a more elegant formulation of anomaly
cancellation can be found in \cite{dmf}.
Second, we have modified some of the bosonic terms in the supergravity
Lagrangian but supersymmetry will clearly require modifications to
fermion terms as well. These have not been worked out to my knowledge.

\subsection{Applications of M5-brane anomaly cancellation}

The cancellation of anomalies for the M theory fivebrane has some
interesting applications. These will be described only briefly here.
For further details see \cite{hmm,Intriligator,Berman,KrausLarsen}.

Earlier we used the known zero modes of the charge $Q_5=1$ 
M5-brane to deduce
the bulk coupling $\int C_3 X_8$ given in \calle{baf}. In the previous
section we have understood all the bulk contributions necessary to
cancel both the tangent and normal bundle anomalies, again for
$Q_5=1$. For $Q_5>1$ we do not know very much about the $(2,0)$
supersymmetric theory of the
zero modes of the M5-brane. In particular we do not know how to
compute the zero mode contribution to the tangent and normal bundle
anomalies. However we do understand the anomaly inflow of the bulk
couplings for $Q_5>1$ and so we can use these to predict what the
zero mode contributions must be, assuming that the anomalies do
indeed cancel.

For a charge $Q_5$ fivebrane $G_4$ obeys the equation
\beq
\label{bauu}
d G_4 = 2 \pi Q_5 d \rho \wedge e_4/2
\eeq
that is, $G_4$ scales linearly with $Q_5$. The two bulk terms which
contribute to the anomaly are the $\int C_3 X_8$ term which scales
like $Q_5$ and the Chern-Simons term which scales like $Q_5^3$. Anomaly
cancellation therefore predicts that the M5-brane zero mode contribution
to the anomaly should be
\beq
\label{batt}
I_8^{zm}(Q_5) = Q_5 I_8^{zm}(1) + (Q_5^3 - Q_5) {p_2(N) \over 24}
\eeq
with
\beq
\label{bass}
I_8^{zm}(1) = {1 \over 48} \left[ p_2(N) - p_2(TW^6) +
{1 \over 4} (p_1(TW^6) - p_1(N))^2 \right] 
\eeq

In the $(2,0)$ theory the $SO(5)$ R symmetry current, whose anomaly we have
just deduced, is in the same supermultiplet as the energy-momentum tensor.
So in principle the anomaly \calle{batt} has interesting implications for various
correlation functions of the $(2,0)$ theory. In particular, the $Q_5^3$
dependence in \calle{batt} is consistent with the $Q_5^3$ dependence found in
the entropy of black holes related to the $(2,0)$ theory \cite{Klebanov,Gubser} and
calculations of the conformal anomaly using the AdS/CFT correspondence \cite{Henningson}.

We have assumed above that the theory is at the origin of moduli space, with all $Q_5$
M5-branes coincident and the $SO(5)_R$ symmetry unbroken. As we move away from
the origin by separating the M5-branes we should integrate out the fields which become
massive to derive the low-energy effective theory of separated fivebranes. This low-energy theory
naively consists of $Q_5$ free $(2,0)$ theories. However, as is well known from analogous
considerations in the analysis of chiral symmetries in QCD, the story is more complicated.
Wess-Zumino terms are generated when we integrate out fermions which contribute to the
anomaly, and these WZ terms have an anomalous variation which ensures that the anomaly
matches throughout moduli space. These WZ terms for the $(2,0)$ theory were worked out
in \cite{Intriligator} and have some interesting implications for the structure of $(2,0)$ theories.

The $(2,0)$ theory is known to have self-dual string excitations \cite{CallanMaldacena,Gibbons,HoweLambertWest}. From the spacetime point of view, these arise as the boundaries of M2-branes
ending on M5-branes. These self-dual strings carry fermion zero modes with a non-zero anomaly,
and techniques similar to those described here can be used to deduce information about the scaling of
these anomalies \cite{Berman} with $Q_5$ and $Q_2$, the number of M2-branes or equivalently, the
charge of the self-dual string.

In \cite{msw} a model of black holes in theories with $N=2$ supersymmetry
was introduced which uses the properties of M theory fivebranes. One
considers a compactification of M theory on a Calabi-Yau space $X$ and
considers a M5-brane which wraps a supersymmetric four-cycle
$P_4 \subset X$. In the resulting five dimensional theory the wrapped
fivebrane appears as a one-brane or string. If one takes the direction
along the string to be a circle and gives the string non-zero
momentum $q$ along the string, then the description of this
configuration in $D=4$ is that of a extremal black hole with
non-zero entropy.  

At low-energies the effective
theory on the string is given by a $(4,0)$ SCFT, and the entropy of the
black hole in this model is determined by the left-moving conformal
anomaly $c_L$ and the charge $q$. In \cite{msw}\ $c_L$ was determined by an
intricate index theory computation. It can however be determined purely
by anomaly inflow. The $(4,0)$ SCFT has a $SU(2)_R$ affine Lie algebra
related to
the normal bundle of the string.  The level $k$ of this $SU(2)_R$ is
determined directly by the normal bundle anomaly and is related by
supersymmetry to the right-moving central charge $c_R$. On the other
hand cancellation of the tangent bundle anomaly of the string determines
$c_R - c_L$ so $c_L$ can be computed purely by anomaly inflow. Recently, similar
techniques have been applied to explain spacetime corrections to black hole
entropy coming from higher order terms in the spacetime effective action \cite{KrausLarsen}.

\subsection{Exercises for Lecture 5.}

\begin{itemize}
\item Show that $\alpha_{n-1} = \epsilon_{a_1\cdots a_n} dy^{a_1} \cdots
dy^{a_{n-1}} y^{a_n}$ is proportional to the volume form on the $(n-1)$ sphere with
the $y^{a_i}$ being Cartesian coordinates on $R^n$.
Hint:  Use Stokes theorem and the fact that $d \alpha_{n-1}$ is proportional to the
volume form on $R^n$.
\item Compute the contribution to the five-brane anomaly from the interaction
\calle{bae} and verify that the tangent bundle anomaly from this term cancels the
tangent bundle anomaly from the five-brane zero modes.
\item Following the treatment in section 5.3, construct a smoothed out source
for a one-brane in four dimensions, $\delta_2(\Sigma^2 \hookrightarrow M^4)$
and verify  that $\delta_2|_{\Sigma^2} = e(F)$, a result which was used in the
analysis of normal bundle anomaly cancellation for axion strings. 
\end{itemize}

\section*{Acknowledgements}

I would like to thank various colleagues and collaborators for many
useful discussions on anomalies. These include D.~Berman,
R.~Bott, A.~Boyarsky, J.~Blum, C.~Callan, L.~Dixon,  D.~Freed, P.~Ginsparg, M.~Green, D.~Kutasov, E.~Martinec, R.~Minasian,
G.~Moore, S.~Naculich, O.~Ruchayskiy, and  E.~Witten. Special thanks
go to S.~Jensen for a careful reading of the manuscript and many constructive criticisms. 
I also thank the student participants
of TASI99, TASI2003  and the Hangzhou Spring School in 2005 as well as the
particle theory graduate students at the University of Chicago for their
encouragement and for many useful questions and discussions on anomalies. This work was supported in part by NSF Grant No. PHY-0204608.

\vfil \eject

\end{document}